\documentclass[aps,prl,twocolumn,superscriptaddress,showpacs,floatfix]{revtex4-1}
\usepackage{graphicx}
\usepackage{physics}
\usepackage{amsmath}
\usepackage{wasysym}
\usepackage{graphicx}
\usepackage{gensymb}
\usepackage{braket}
\usepackage{natbib}
\usepackage{comment}
\usepackage{physics}
\usepackage{color}
\usepackage[normalem]{ulem}
\usepackage[colorlinks=true,citecolor=blue,linkcolor=magenta, breaklinks=true]{hyperref}

\begin{document}
	
	\title{Emergent Quadrupolar Order in the Spin-$1/2$ Kitaev-Heisenberg Model}
	
	\begin{abstract}
		Motivated by the largely unexplored domain of multi-polar ordered spin states in the Kitaev-Heisenberg (KH) systems we investigate the ground state dynamics of the spin-$\frac{1}{2}$ KH model, focusing on quadrupolar (QP) order in 2-leg ladder and two-dimensional honeycomb lattice geometries. Employing exact diagonalization and density-matrix renormalization group methods, we analyze the QP order parameter and correlation functions. Our findings reveal a robust QP order across a wide range of the phase diagram, influenced by the interplay between Heisenberg and Kitaev interactions. Notably, we observe an enhancement of QP order near Kitaev quantum spin liquid (QSL) phases, despite the absence of long-range spin-spin correlations. This highlights a complex relationship between QP order and QSLs, offering new insights into quantum magnetism in low-dimensional systems. Our findings provide a rational explanation for the observed nonlinear magnetic susceptibility in $\alpha$-RuCl$_3$.
	\end{abstract}
	
	
	\author{Manodip Routh}
	\thanks{These authors contributed equally to this work.}
	\affiliation{S.N. Bose National Centre for Basic Sciences, Kolkata 700098, India.}
	\author{Sayan Ghosh}
	\thanks{These authors contributed equally to this work.}
	\affiliation{S.N. Bose National Centre for Basic Sciences, Kolkata 700098, India.}
	\author{Jeroen van den Brink}
	\affiliation{Institute for Theoretical Solid State Physics, IFW Dresden, 01069 Dresden, Germany.}
	\affiliation{\mbox{W\"urzburg-Dresden Cluster of Excellence ct.qmat, Germany}}
	\affiliation{Department of Physics, Technical University Dresden, 01069 Dresden, Germany.}
	\author{Satoshi Nishimoto}
	\email{s.nishimoto@ifw-dresden.de}
	\affiliation{Institute for Theoretical Solid State Physics, IFW Dresden, 01069 Dresden, Germany.}
	\affiliation{Department of Physics, Technical University Dresden, 01069 Dresden, Germany.}
	\author{Manoranjan Kumar}
	\email{manoranjan.kumar@bose.res.in}
	\affiliation{S.N. Bose National Centre for Basic Sciences, Kolkata 700098, India.}
	\date{\today}
	
	\maketitle	
	\textit{Introduction}---
	Quantum spin liquids (QSLs) represent a quantum phase of matter that challenges traditional understanding by lacking magnetic long-range order (LRO), yet they may exhibit global topological order
	\cite{balents2010spin}. These elusive states emerge in frustrated magnetic systems, where competing spin exchange interactions not only foster the QSL state but also lead to complex entangled states such as dimer formations \cite{chhajlany2007entanglement}, fractional excitations like spinons \cite{enderle2010two,schlappa2018probing}, and Majorana excitations \cite{do2017incarnation,koga2021majorana,wulferding2020magnon,hashimoto2022majorana} in the ground state (gs). A key development in understanding QSLs was the Kitaev model, which proposed spin-$1/2$ particles on a hexagonal lattice with direction-dependent exchange interactions, introducing frustration into the system \cite{KITAEV20062}. The gs of this model is a QSL, characterized by Majorana fermions and gauge vortices as elementary excitations \cite{hermanns2015weyl,le2017majorana,mandal2011confinement,2007Baskaran,2008Baskaran, baskaran2023metastable}.
	
	The experimental pursuit of the Kitaev model has prominently featured Iridium-based materials such as A$_2$IrO$_3$ (A=Na, Li), where Ir atoms form a honeycomb structure \cite{singh2012relevance, hwan2015direct}. In these materials, the octahedral coordination of ligands induces a crystal field, splitting the Ir $d$-orbitals into $t_{2g}$ and $e_g$ levels. Strong spin-orbit coupling further splits the $t_{2g}$ orbitals, allowing the two highest states of each to mimic effective spin-$1/2$ degrees of freedom. This results in orbital-dependent, directionally anisotropic spin exchanges, i.e., so-called Kitaev interactions \cite{TREBST20221}. However, the Kitaev-Heisenberg (KH) model, incorporating residual nearest-neighbor Heisenberg interactions, offers a more realistic description of the magnetic properties of these materials \cite{2010PRLIrridium1}. Similar Hamiltonians have been proposed for materials like $\alpha$-RuCl$_3$ \cite{suzuki2021proximate,do2017majorana,banerjee2017neutron,trebst2022kitaev,winter2016challenges,janssen2016honeycomb}, $\beta$-Li$_2$IrO$_3$ \cite{katukuri2016vicinity,takayama2015hyperhoneycomb,trebst2022kitaev,winter2016challenges,janssen2016honeycomb}, and $\gamma$-Li$_2$IrO$_3$ \cite{tsirlin2022kitaev,trebst2022kitaev}.
	
	The gs phase diagram of the honeycomb-lattice KH model has been extensively studied, revealing four magnetic LRO phases (N\'eel, zigzag (ZZ), stripy (ST), and ferromagnetic (FM)) and two QSL phases; FM Kitaev (FK) and antiferromagnetic Kitaev (AFK) QSLs by tuning the Heisenberg and Kitaev interactions \cite{2010PRLIrridium1,gotfryd2017phase,schaffer2012quantum,feng2007topological}. However, the potential for a quantum phase involving multi-polar spin states, such as a quadrupolar (QP) or spin-nematic state, in the KH model remains largely unexplored. From an experimental perspective, third-order positive susceptibility probing a higher-order correlation such as QP has been observed at zero magnetic field in $\alpha$-RuCl$_3$ \cite{holleis2021anomalous}. Yet, the intricate mechanisms behind these observations are not fully elucidated. In addition to these experimental findings, the present study is further motivated by the inclusion of a staggered QP operator in the Hamiltonian of the KH model (see below).
	
	The QP phase, established in liquid crystal systems \cite{1973kelker,1988kelker,1966MerminWagner,2012liquidcrystalRevModPhys}, manifests in magnetic systems as spin nematics, where magnetic quadrupole moments create orientational order without magnetic LRO \cite{2006lauchili, 2006Tsunetsugu, lacroix2011introduction}. Initially identified in spin-1 systems with biquadratic exchange \cite{blume1969biquadratic}, this order has been explored in various Heisenberg spin-1 models \cite{hu2019nematic,podolsky2005properties,tanaka2020multiple}. Extending this concept to spin-$1/2$ systems has spurred research, particularly in $J_1$$-$$J_2$ chain systems under strong magnetic fields \cite{hikihara2008vector,parvej2017multipolar,sudan2009emergent}. In the context of Kitaev systems, a spin-nematic phase in the spin-$1/2$ Kitaev-Ising model has been predicted, arising from the interplay between Kitaev QSL and magnetic LRO, but lacking topological order \cite{2017PRLNasu}. Additionally, four-body interactions among Majorana fermions in the Kitaev QSL could induce a topological nematic phase transition, evolving from the chiral QSL phase to the toric code phase \cite{2021PRRTakahashi}.
	
	In this letter, we explore the gs dynamics of the spin-$\frac{1}{2}$ KH model, focusing on the QP order. We investigate two geometries: the 2-leg ladder and the two-dimensional (2D) honeycomb lattice, using exact diagonalization (ED) and density-matrix renormalization group (DMRG) methods. Our primary contribution is a detailed exploration of the QP order parameter and correlation functions within these systems. We present a novel gs phase diagram highlighting the QP order as a function of the ratio between Heisenberg and Kitaev interactions. A striking aspect of our findings is the robust presence and stability of QP order across a wide range of the phase diagram, with a notable enhancement near the Kitaev QSL phases. This enhancement occurs despite the absence of longer-range spin-spin correlations, adding a new dimension to our understanding of the interplay between QP order and QSLs. 
	
	\textit{Quadrupolar operator}---
	The QP state emerges as a quantum ordered phase characterized by collective bimagnon excitations. This intriguing state, while exhibiting a preference for the orientation of paired magnons, notably does not break time-reversal symmetry. Its detection hinges on the analysis of a symmetric and traceless rank-2 tensor operator \cite{Penc2011}:
	\begin{align}
		\hat{Q}_{ij}^{\alpha\beta} = S_i^{\alpha}S_j^{\beta} + S_j^{\alpha}S_i^{\beta} - \frac{2}{3}(S_i.S_j)\delta_{\alpha\beta}
		\label{Eq:Qab}
	\end{align}
	where $\alpha$ and $\beta$ represent the Cartesian coordinates such as $x$, $y$, and $z$. As explained below, the Hamiltonian of the KH model can be reformulated to explicitly include a specific component of Eq. \eqref{Eq:Qab}:
	\begin{align}
		\hat{Q}^{x^2-y^2}_{ij}=\frac{1}{2}(S^+_{i,j} S^+_{i+1,j} + S^-_{i,j} S^-_{i+1,j}).
		\label{Eq:Qx2y2}
	\end{align}
	
	\begin{figure}[t]
		\centering
		\includegraphics[width=1.0\columnwidth]{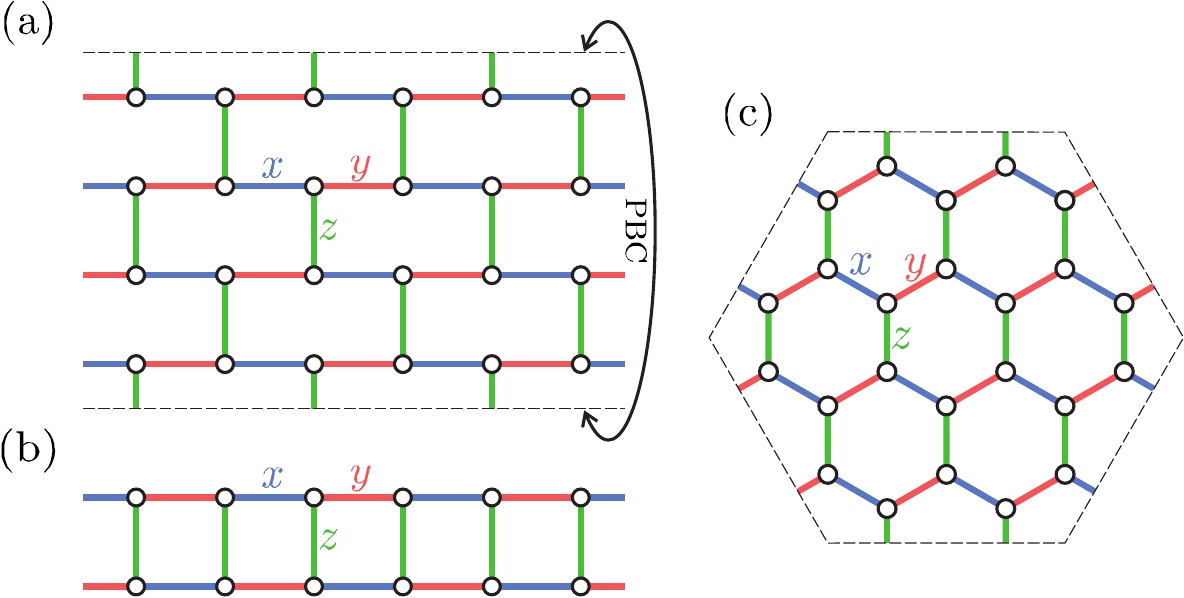} 
		\caption{Lattice structures used in our calculations: (a) KH ladder or brickwall lattice which is obtained by deforming the honeycomb-lattice KH model. (b) 2-leg KH ladder. (c) 24-site honeycomb-lattice KH cluster with periodic boundary conditions.
		}
		\label{fig:Figure1}
	\end{figure}
	
	\textit{Model}---
	We consider a spin-$1/2$ KH model on both a ladder and a honeycomb lattice \cite{2010PRLIrridium1}. The Hamiltonian is given by
	\begin{align}
		\mathcal{H}=J\sum_{i,j}\vec{S}_i\cdot\vec{S}_j + K\sum_{\gamma} \sum_{i,j} S_i^\gamma S_j^\gamma.
		\label{Eq:KH_eq}
	\end{align}
	Where $S_i^\gamma$ is the $\gamma$ ($=x$, $y$, or $z$) component of the spin-$1/2$ operator $\vec{S}_i$ at site $i$. Here, $J$ and $K$ denote the Heisenberg and Kitaev interactions, respectively. To facilitate analysis, we introduce an angular parameter $\phi$ ($\in[0, 2\pi]$), defining $J=\cos \phi$ and $K=\sin \phi$.
	
	For the ladder or cylinder geometry, the Hamiltonian is adapted to include the spin-$1/2$ operator $\vec{S}{i,j}$ at site $i$ on the $j$-th leg:
	\begin{align}
		&\mathcal{H}_{\rm leg} = \frac{2J+K}{4}\sum_{j=1}^{n}\sum_{i=1}^{L}(S_{i,j}^{+}S_{i+1,j}^{-}+S_{i,j}^{-}S_{i+1,j}^{+}) \nonumber \\
		& +\frac{K}{2}\sum_{j=1}^{n}\sum_{i=1}^{L}(-1)^{i+j}(\hat{Q}^{x^2-y^2}_{ij})
		+J \sum_{j=1}^{n}\sum_{i=1}^{L}(S_{i,j}^zS_{i+1,j}^z) \nonumber \\
		&\equiv \mathcal{H}_{\rm exc}+\mathcal{H}_{\rm Q}+\mathcal{H}_{\rm z}.
		\label{Eq:KH_eq_leg}
	\end{align}
	In this formulation, $\mathcal{H}_{\rm Q}$ signifies the QP operator. Its dominance suggests the potential stabilization of QP order. The possibility of QP order arising through similar mechanisms has been previously discussed in the context of $J_1$-$J_2$ chain systems under magnetic fields \cite{grafe2017signatures}.
	
	Although the Kitaev model was originally proposed for the honeycomb lattice, it is recognized that any three-coordinated lattice with Kitaev interactions can exhibit similar intriguing properties. In this sense, the KH model on a 2-leg ladder [Fig. \ref{fig:Figure1}(b)], derived from a brick-wall lattice analogous to the honeycomb lattice, aligns with the necessary geometry. This alignment indicates that the 2-leg KH ladder can offer vital insights into the behavior of the 2D honeycomb-lattice KH model. Notably, the gs phase diagram of the ladder model displays significant similarities with the 2D counterpart in terms of magnetic order and Kitaev QSL, with the key distinction being the replacement of the N\'eel phase by the rung-singlet (RS) phase \cite{agrapidis2019ground,catuneanu2019nonlocal}. Therefore, initiating our investigation with the 2-leg KH ladder is a strategic choice. This setup not only allows for precise numerical calculations but also facilitates reliable extrapolations to the thermodynamic limit, essential for exploring novel phenomena in the 2D KH model. Building upon this, we extend our investigation to 4-leg and 6-leg cylinders [Fig. \ref{fig:Figure1}(a)], progressively approaching the 2D bulk limit. This systematic increase in the number of legs serves to bridge our understanding from the simpler ladder systems to the more complex dynamics inherent in the full 2D KH model.
	
	\textit{Methods}---
	To investigate the intricate dynamics of the KH model, we employ two complementary computational techniques. For the 2-leg ladder system [Fig. \ref{fig:Figure1}(b)], encompassing system sizes up to $12 \times 2$, and for the 24-site periodic boundary condition (PBC) cluster [Fig. \ref{fig:Figure1}(c)], we utilize the exact diagonalization (ED) method. For larger systems, specifically extended 2-leg ladders, as well as 4-leg and 6-leg cylinders [Fig. \ref{fig:Figure1}(a)], we implement the DMRG method. In our DMRG calculations, we retain up to 5000 density-matrix eigenstates during the renormalization process. This high number of kept states ensures the accuracy of our results, with the largest discarded weight being maintained at an impressively low level of approximately $2 \times 10^{-5}$.
	
	\begin{figure}[t]
		\includegraphics[width=1.0\columnwidth]{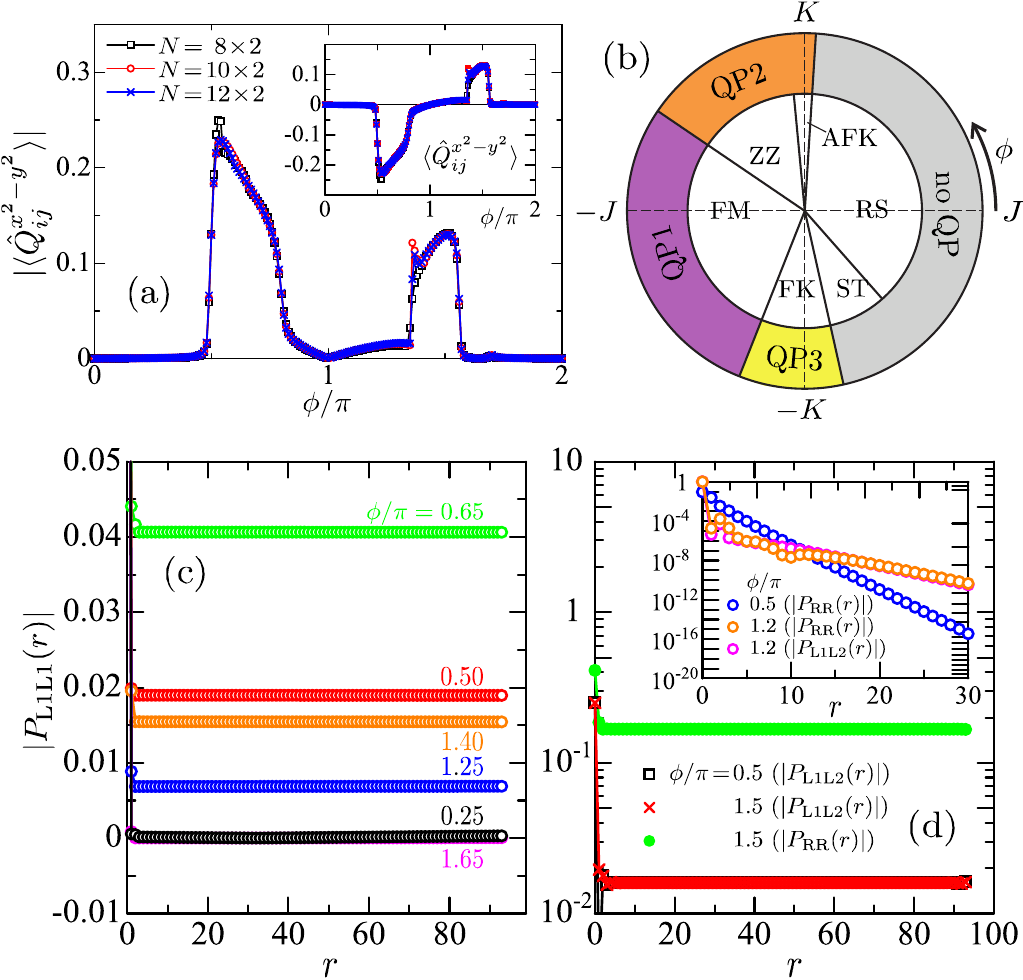}
		\caption{Results for 2-leg KH ladder. (a) Absolute values of the QP order parameter for leg bonds with various system sizes. The inset represents the same physical quantity but without taking absolute values. (b) Ground-state Phase diagram for magnetic order (inner) and QP order (outer). Four long-range ordered phases: rung-singlet (RS), zigzag (ZZ), ferromagnetic (FM), stripy (ST) and two QSL phases: antiferromagnetic Kitaev (AFK) and ferromagnetic Kitaev (FK). (c) Compilation of the behavior of the QP correlation functions $|P_{\rm L1L1}(r)|$ with $N=100 \times 2$ cluster for representative parameters (see text). (d) Inter-leg and rung-rung QP correlation functions at $\phi/\pi=0.5$, $1.2$, and $1.5$. Inset: Examples for the cases where the QP correlation function exhibits an exponential decay.}
		\label{fig:twoleg}
	\end{figure}
	
	\textit{Quadrupolar order in the 2-leg ladder}---
	We commence our analysis with the 2-leg ladder system. To obtain a comprehensive phase diagram for QP order as a function of $\phi$, we calculate the QP order parameter $\langle \hat{Q}^{x^2-y^2}_{ij} \rangle$ using ED. Our findings, focusing on a single bond along the leg, are depicted in Fig.\ref{fig:twoleg}(a). We find that $\langle \hat{Q}^{x^2-y^2}_{ij} \rangle$ exhibits finite values over a broad range of $\phi$ ($0.48\pi \lesssim \phi \lesssim 1.56\pi$) with minimal finite-size effects. This order parameter parameter shows significant values in the AFK, FK, and ZZ phases of the magnetic phase diagram, while it rather diminishes in the FM phase [Fig.\ref{fig:twoleg}(b)]. Conversely, when $i$ and $j$ are positioned on a rung bond, $\langle \hat{Q}^{x^2-y^2}_{ij} \rangle$ consistently equals zero.
	
	To further substantiate these findings, using DMRG we compute leg-leg and rung-rung QP correlation functions, defined as:
	\begin{align}
		P_{\rm LmLn}(r) = \langle S_{i,m}^+ S_{i+1,m}^+ S_{i+r,n}^- S_{i+1+r,n}^- \rangle
		\label{Eq:PLmLn}
	\end{align}
	and
	\begin{align}
		P_{\rm RR}(r) = \langle S_{i,m}^+ S_{i,m+1}^+ S_{i+r,m}^- S_{i+r,m+1}^- \rangle,
		\label{Eq:PRR}
	\end{align}
	respectively. The results, summarized in Fig.\ref{fig:twoleg}(c,d), reveal that the intra-leg correlation $P_{\rm L1L1}(r)$ exhibits QP LRO within the $\phi$ range where $\langle \hat{Q}^{x^2-y^2}_{ij} \rangle$ is finite. Note that $P_{\rm L1L1}(r)$ alternates in sign with distance. Meanwhile, near both Kitaev points $\phi=\pm \pi/2$, the inter-leg correlation $P_{\rm L1L2}(r)$, and near the FK phase, the rung-rung correlation $P_{\rm RR}(r)$, also exhibit QP LRO. Unlike $P_{\rm L1L1}(r)$, both $P_{\rm L1L2}(r)$ and $P_{\rm RR}(r)$ are always positive. Interestingly, despite the rung QP order parameter being consistently zero across all $\phi$ values, $P_{\rm RR}(r)$ still exhibits LRO in the FK phase. This suggests that states of QP order with broken spin-rotational symmetry are degenerate, leading to a zero local order parameter, yet detectable through correlation functions.
	
	From these correlation function analyses, the $\phi$ domain for QP order can be classified into four phases: QP1 (only $P_{\rm L1L1}(r)$ shows LRO), QP2 (both $P_{\rm L1L1}(r)$ and $P_{\rm L1L2}(r)$ show LRO), QP3 (all of $P_{\rm L1L1}(r)$, $P_{\rm L1L2}(r)$, and $P_{\rm RR}(r)$ show LRO), and a phase with no QP LRO. The schematic picture of each QP phase is illustrated in the Supplementary Material \cite{suppmat}. The correspondence with the magnetic order phases is shown in the phase diagram in Fig.\ref{fig:twoleg}(b). The phase diagram for QP order can be reasonably explained by considering the spin structures of the corresponding magnetic phases. In the FM phase ($0.81 \lesssim \phi/\pi \lesssim 1.38$), the near-full magnetization results in smaller values of $P_{\rm L1L1}(r)$ and $|\langle \hat{Q}^{x^2-y^2}_{ij} \rangle|$, and they vanish without $\mathcal{H}_{\rm Q}$ at $\phi=\pi$. The ZZ phase ($0.53 \lesssim \phi/\pi \lesssim 0.81$) exhibits Ising-type LRO in each leg with opposite polarization between legs \cite{agrapidis2019ground}, leading to the anticipated development of $P_{\rm L1L2}(r)$. However, $P_{\rm RR}(r)$ does not show LRO due to antiparallel spin orientation on the rungs. In fact, a detailed wave-function analysis suggests the dominant contribution of double-flip configurations in the gs (see the Supplementary Material \cite{suppmat}). In the RS ($-0.26 \lesssim \phi/\pi \lesssim 0.48$) and ST ($1.57 \lesssim \phi/\pi \lesssim 1.74$) phases, the each leg possesses a N\'eel-type LRO, precluding the dominance of the $\mathcal{H}_{\rm Q}$ term. Thus, a QP order is absent.
	
	In the AFK ($0.48 \lesssim \phi/\pi \lesssim 0.53$) and FK ($1.38 \lesssim \phi/\pi \lesssim 1.57$) phases, the competition between dipole fluctuations ($\mathcal{H}_{\rm exc}$) and QP fluctuations ($\mathcal{H}_{\rm Q}$) complicates gs determination. Nevertheless, the larger coefficient of $\mathcal{H}_{\rm Q}$ supports the emergence of QP order along the leg, as corroborated by our numerical results. Remarkably, in the FK phase, the rung QP correlation function $P_{\rm RR}(r)$ also exhibits LRO, with a saturation value exceeding $P_{\rm L1L1}(r)$. This is attributed to the FM Ising character of rung interactions, conducive to bimagnon formation \textcolor{blue}{\cite{Shannon2006}}. The approximate wave function of the FK phase is detailed in the Supplementary Material \cite{suppmat}.
	
	Given these insights, we hypothesize that QP order might also be present in the 2D honeycomb KH system, akin to the 2-leg ladder scenario. To test this hypothesis, we extend our investigation to isotropic honeycomb cluster as well as 4-leg and 6-leg cylinders, in the parts that follow.
	
	\begin{figure}[t]
		\centering
		\includegraphics[width=1.0\columnwidth]{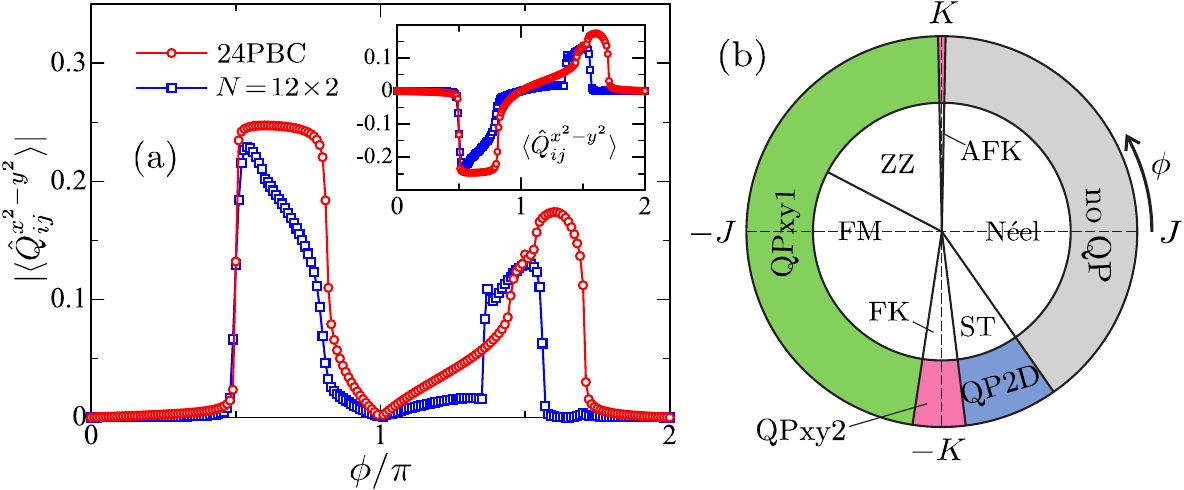} 
		\caption{Results for honeycomb-lattice KH model. (a) Absolute values of the QP order parameter for $x$ or $y$ bond, calculated using the 24-site PBC cluster. For comparison, those for the 2-leg KH ladder are also plotted. The inset represents the same physical quantity but without taking absolute values. (b) Ground-state Phase diagram for magnetic order (inner) and QP order (outer).} 
		\label{fig:quad_honeycomb}
	\end{figure}
	
	\textit{Quadrupolar order in the 2D honeycomb lattice}---
	Analogous to the approach for the 2-leg ladder, we first aim to grasp the overall picture of the QP order phase diagram as a function of $\phi$ by calculating the QP order parameter using ED on the 24-site PBC cluster.The results, illustrated in Fig. \ref{fig:quad_honeycomb}(a), are compared with those from the 2-leg ladder. In the context of magnetic order phases for the 2D honeycomb-lattice KH model, the QP order parameter is finite in the AFK ($0.494 \lesssim \phi/\pi \lesssim 0.506$), ZZ ($0.506 \lesssim \phi/\pi \lesssim 0.847$), FM ($0.847 \lesssim \phi/\pi \lesssim 1.452$), FK ($1.452 \lesssim \phi/\pi \lesssim 1.539$), and ST ($1.539 \lesssim \phi/\pi \lesssim 1.694$) phases, and it is zero only in the N\'eel phase ($-0.306 \lesssim \phi/\pi \lesssim 0.494$). While the results are fundamentally similar to those for the 2-leg ladder, a notable difference is that the QP order parameter remains finite even in the ST phase. For the $z$-bonds, similar to the rung QP in the 2-leg ladder, the QP order parameter is zero. 
	
	\begin{figure}[t]
		\centering
		\includegraphics[width=1.0\columnwidth]{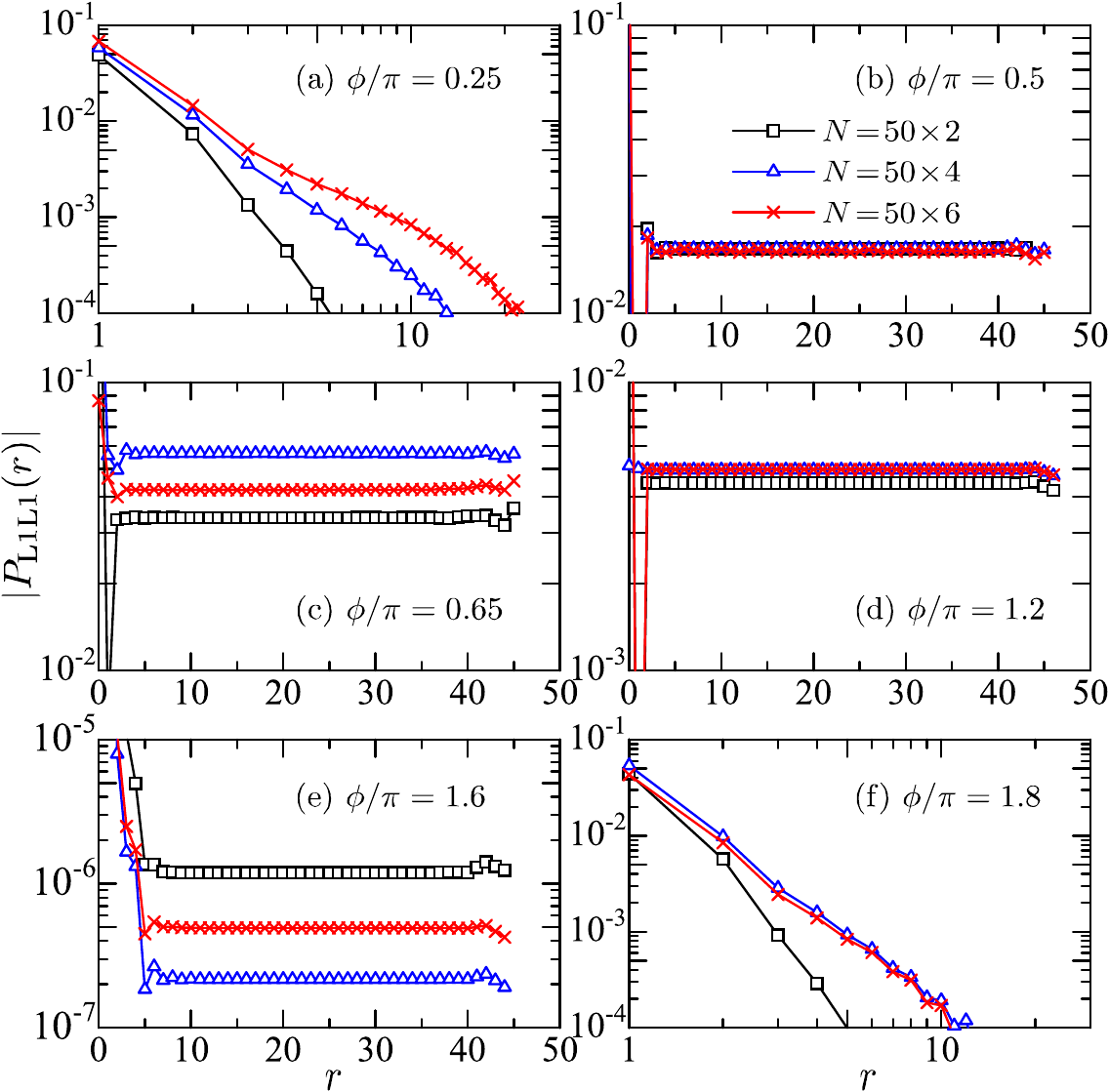} 
		\caption{DMRG results for the QP correlation functions $P_{\rm L1L1}(r)$ at representative value of $\phi$ for six quantum phases using 2-leg, 4-leg, and 6-leg cylinders.
		} 
		\label{fig:2DPL1L1}
	\end{figure}
	
	To ascertain the persistence of QP order in the bulk limit, we compute QP correlation functions using DMRG for 4-leg and 6-leg cylinders, in addition to the 2-leg ladder. Focusing on the intra-leg QP correlation function, $P_{\rm L1L1}(r)$, we observe in Fig. \ref{fig:2DPL1L1} that, consistent with the QP order parameter results, $P_{\rm L1L1}(r)$ converges to a finite value in the long-distance limit for all magnetic phases except the N\'eel phase. Importantly, this convergence value does not diminish with an increasing number of legs, suggesting its stability in the bulk limit. In the N\'eel phase, $P_{\rm L1L1}(r)$ exhibits a power-law decay indicating no long-range QP order.
	
	\begin{figure}[t]
		\centering
		\includegraphics[width=1.0\columnwidth]{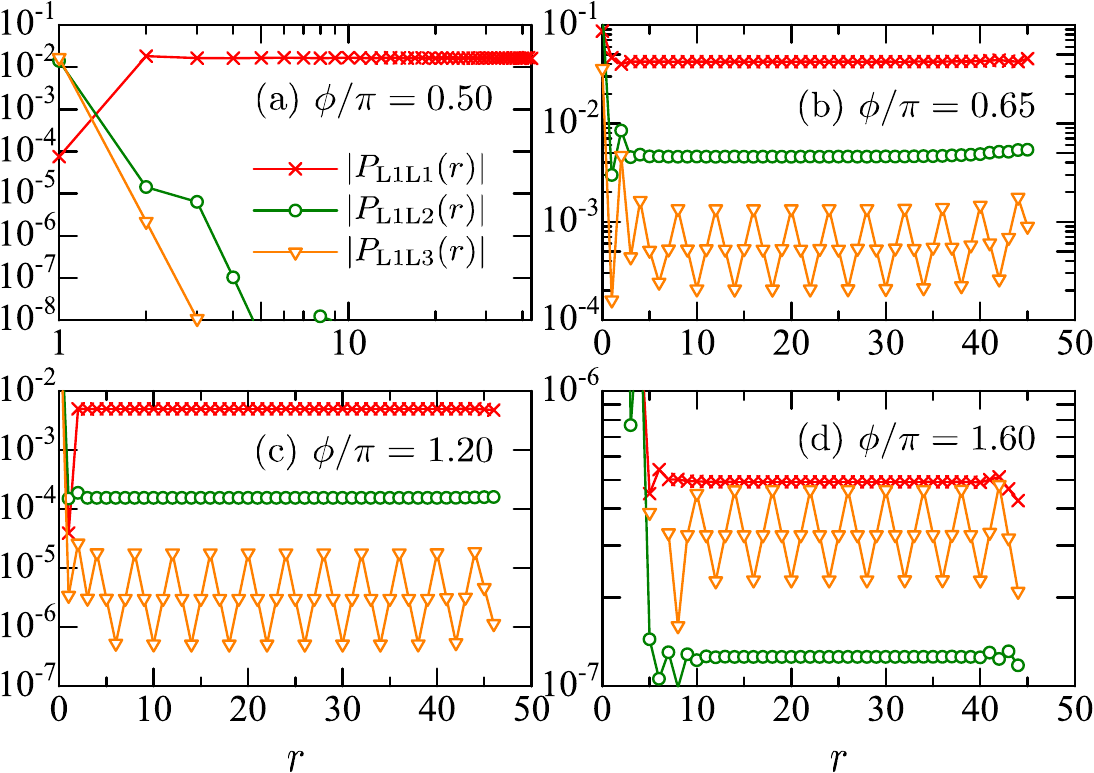} 
		\caption{DMRG results for the inter-leg QP correlation functions, $P_{\rm L1Ln}(r)$ for $n=1 - 3$ at magnetic ordered phases, including ZZ ($\phi/\pi = 0.65$), FM ($\phi/\pi = 1.2$), ST ($\phi/\pi = 1.6$) and QSL phase ($\phi/\pi = 0.5$), using 6-leg cylinders.
		} 
		\label{fig:2DPL1Ln}
	\end{figure}
	
	The decay of QP correlation functions perpendicular to the leg direction, namely along the $z$-bond, is also of interest. We calculate the inter-leg QP correlation function, $P_{\rm L1Ln}(r)$, for various leg distances in a $50 \times 6$ system. The calculated $P_{\rm L1Ln}(r)$ for $n=1-3$ is plotted in Fig. \ref{fig:2DPL1Ln}. In the ZZ and FM phases, the saturation value of $P_{\rm L1Ln}(r)$ at long distances follows power-law decay [Fig.~\ref{fig:2DPL1Ln}(b,c)], suggesting quasi-LRO along the $z$-bond direction (QPxy1 phase). Near the Kitaev points (AFK and FK), a rapid decay of inter-leg correlations is observed [Fig.~\ref{fig:2DPL1Ln}(a)], indicating LRO in the $xy$ zigzag direction alone (QPxy2 phase). The ST phase presents a unique case, where the decay along the $z$-bond direction does not follow a simple pattern [Fig.~\ref{fig:2DPL1Ln}(d)], hinting at the potential for LRO (QP2D phase). For all $\phi$ values, the rung-rung QP correlation function, $P_{\rm RR}(r)$, exhibits exponential decay for both 4-leg and 6-leg cylinders. The correspondence between QP order and magnetic order is summarized in the phase diagram in Fig. \ref{fig:quad_honeycomb}(b). A notable distinction from the 2-leg ladder case is the presence of QP order in the ST phase, likely due to the unique orientation of ST order in this setting, although the magnitude of QP correlations in this phase is relatively small compared to other phases.

	\textit{Summary}---
	We have explored the gs dynamics of the spin-$\frac{1}{2}$ KH model, focusing particularly on the emergence and characteristics of QP order. Our study spans two distinct geometries: the 2-leg ladder and the 2D honeycomb lattice. Employing a synergistic approach combining ED and DMRG techniques, we have meticulously analyzed the QP order parameter and correlation functions in these systems. Our key contribution is the unveiling of a comprehensive gs phase diagram, which distinctly maps out the QP order in relation to the interplay between Heisenberg and Kitaev interactions. A notable discovery is the pronounced stability of QP order across a broad range of the phase diagram, underscoring its robustness in the KH model. Additionally, our findings reveal a remarkable enhancement of the QP order parameter and its correlation functions near the Kitaev QSL phases. This enhancement is particularly intriguing as it occurs in the absence of long-range spin-spin correlations, suggesting a nuanced relationship between QP order and QSLs. In fact, the third-order positive susceptibility, indicative of a general signature of QP order, has been observed in $\alpha$-RuCl$_3$ \cite{holleis2021anomalous}, where the relationship between $K$ and $J$ appears to be $-K \gg J > 0$ \cite{yadav2016kitaev}, suggesting a proximity to the FK QSL phase.
	
	These insights not only advance our understanding of the KH model but also contribute significantly to the broader discourse on quantum magnetism in low-dimensional systems, highlighting the intricate interplay between different types of quantum order.
	
	{\bf Acknowledgments}$-$ MK thanks Saptarshi Mandal, Krishnendu Sengupta, David Logan, Jay D Sau, Rajiv R. P. Singh for the fruitful discussions. We thank Ulrike Nitzsche for technical assistance. MK acknowledges support from SERB through Grant Sanction No. CRG/2020/000754. MR acknowledges the support from Institute for Theoretical Solid State Physics of IFW Dresden and DST-India. SG expresses appreciation for financial support from DST-INSPIRE.  This project is funded by the German Research Foundation (DFG) via the projects A05 of the Collaborative Research Center SFB 1143 (project-id 247310070) and through the Wurzburg-Dresden Cluster of Excellence on Complexity and Topology in Quantum Matter ct.qmat (EXC 2147, Project No. 39085490).
	
	
	\bibliography{KH_ref}

\begin{thebibliography}{55}%
\makeatletter
\providecommand \@ifxundefined [1]{%
 \@ifx{#1\undefined}
}%
\providecommand \@ifnum [1]{%
 \ifnum #1\expandafter \@firstoftwo
 \else \expandafter \@secondoftwo
 \fi
}%
\providecommand \@ifx [1]{%
 \ifx #1\expandafter \@firstoftwo
 \else \expandafter \@secondoftwo
 \fi
}%
\providecommand \natexlab [1]{#1}%
\providecommand \enquote  [1]{``#1''}%
\providecommand \bibnamefont  [1]{#1}%
\providecommand \bibfnamefont [1]{#1}%
\providecommand \citenamefont [1]{#1}%
\providecommand \href@noop [0]{\@secondoftwo}%
\providecommand \href [0]{\begingroup \@sanitize@url \@href}%
\providecommand \@href[1]{\@@startlink{#1}\@@href}%
\providecommand \@@href[1]{\endgroup#1\@@endlink}%
\providecommand \@sanitize@url [0]{\catcode `\\12\catcode `\$12\catcode
  `\&12\catcode `\#12\catcode `\^12\catcode `\_12\catcode `\%12\relax}%
\providecommand \@@startlink[1]{}%
\providecommand \@@endlink[0]{}%
\providecommand \url  [0]{\begingroup\@sanitize@url \@url }%
\providecommand \@url [1]{\endgroup\@href {#1}{\urlprefix }}%
\providecommand \urlprefix  [0]{URL }%
\providecommand \Eprint [0]{\href }%
\providecommand \doibase [0]{http://dx.doi.org/}%
\providecommand \selectlanguage [0]{\@gobble}%
\providecommand \bibinfo  [0]{\@secondoftwo}%
\providecommand \bibfield  [0]{\@secondoftwo}%
\providecommand \translation [1]{[#1]}%
\providecommand \BibitemOpen [0]{}%
\providecommand \bibitemStop [0]{}%
\providecommand \bibitemNoStop [0]{.\EOS\space}%
\providecommand \EOS [0]{\spacefactor3000\relax}%
\providecommand \BibitemShut  [1]{\csname bibitem#1\endcsname}%
\let\auto@bib@innerbib\@empty
\bibitem [{\citenamefont {Balents}(2010)}]{balents2010spin}%
  \BibitemOpen
  \bibfield  {author} {\bibinfo {author} {\bibfnamefont {L.}~\bibnamefont
  {Balents}},\ }\href {\doibase 10.1038/nature08917} {\bibfield  {journal}
  {\bibinfo  {journal} {Nature}\ }\textbf {\bibinfo {volume} {464}},\ \bibinfo
  {pages} {199} (\bibinfo {year} {2010})}\BibitemShut {NoStop}%
\bibitem [{\citenamefont {Chhajlany}\ \emph {et~al.}(2007)\citenamefont
  {Chhajlany}, \citenamefont {Tomczak}, \citenamefont {W{\'o}jcik},\ and\
  \citenamefont {Richter}}]{chhajlany2007entanglement}%
  \BibitemOpen
  \bibfield  {author} {\bibinfo {author} {\bibfnamefont {R.~W.}\ \bibnamefont
  {Chhajlany}}, \bibinfo {author} {\bibfnamefont {P.}~\bibnamefont {Tomczak}},
  \bibinfo {author} {\bibfnamefont {A.}~\bibnamefont {W{\'o}jcik}}, \ and\
  \bibinfo {author} {\bibfnamefont {J.}~\bibnamefont {Richter}},\ }\href
  {\doibase 10.1103/PhysRevA.75.032340} {\bibfield  {journal} {\bibinfo
  {journal} {Physical Review A}\ }\textbf {\bibinfo {volume} {75}},\ \bibinfo
  {pages} {032340} (\bibinfo {year} {2007})}\BibitemShut {NoStop}%
\bibitem [{\citenamefont {Enderle}\ \emph {et~al.}(2010)\citenamefont
  {Enderle}, \citenamefont {F{\aa}k}, \citenamefont {Mikeska}, \citenamefont
  {Kremer}, \citenamefont {Prokofiev},\ and\ \citenamefont
  {Assmus}}]{enderle2010two}%
  \BibitemOpen
  \bibfield  {author} {\bibinfo {author} {\bibfnamefont {M.}~\bibnamefont
  {Enderle}}, \bibinfo {author} {\bibfnamefont {B.}~\bibnamefont {F{\aa}k}},
  \bibinfo {author} {\bibfnamefont {H.-J.}\ \bibnamefont {Mikeska}}, \bibinfo
  {author} {\bibfnamefont {R.}~\bibnamefont {Kremer}}, \bibinfo {author}
  {\bibfnamefont {A.}~\bibnamefont {Prokofiev}}, \ and\ \bibinfo {author}
  {\bibfnamefont {W.}~\bibnamefont {Assmus}},\ }\href {\doibase
  10.1103/PhysRevLett.104.237207} {\bibfield  {journal} {\bibinfo  {journal}
  {Physical review letters}\ }\textbf {\bibinfo {volume} {104}},\ \bibinfo
  {pages} {237207} (\bibinfo {year} {2010})}\BibitemShut {NoStop}%
\bibitem [{\citenamefont {Schlappa}\ \emph {et~al.}(2018)\citenamefont
  {Schlappa}, \citenamefont {Kumar}, \citenamefont {Zhou}, \citenamefont
  {Singh}, \citenamefont {Mourigal}, \citenamefont {Strocov}, \citenamefont
  {Revcolevschi}, \citenamefont {Patthey}, \citenamefont {R{\o}nnow},
  \citenamefont {Johnston} \emph {et~al.}}]{schlappa2018probing}%
  \BibitemOpen
  \bibfield  {author} {\bibinfo {author} {\bibfnamefont {J.}~\bibnamefont
  {Schlappa}}, \bibinfo {author} {\bibfnamefont {U.}~\bibnamefont {Kumar}},
  \bibinfo {author} {\bibfnamefont {K.}~\bibnamefont {Zhou}}, \bibinfo {author}
  {\bibfnamefont {S.}~\bibnamefont {Singh}}, \bibinfo {author} {\bibfnamefont
  {M.}~\bibnamefont {Mourigal}}, \bibinfo {author} {\bibfnamefont
  {V.}~\bibnamefont {Strocov}}, \bibinfo {author} {\bibfnamefont
  {A.}~\bibnamefont {Revcolevschi}}, \bibinfo {author} {\bibfnamefont
  {L.}~\bibnamefont {Patthey}}, \bibinfo {author} {\bibfnamefont
  {H.}~\bibnamefont {R{\o}nnow}}, \bibinfo {author} {\bibfnamefont
  {S.}~\bibnamefont {Johnston}},  \emph {et~al.},\ }\href {\doibase
  10.1038/s41467-018-07838-y} {\bibfield  {journal} {\bibinfo  {journal}
  {Nature Communications}\ }\textbf {\bibinfo {volume} {9}},\ \bibinfo {pages}
  {5394} (\bibinfo {year} {2018})}\BibitemShut {NoStop}%
\bibitem [{\citenamefont {Do}\ \emph {et~al.}(2017{\natexlab{a}})\citenamefont
  {Do}, \citenamefont {Park}, \citenamefont {Yoshitake}, \citenamefont {Nasu},
  \citenamefont {Motome}, \citenamefont {Kwon}, \citenamefont {Adroja},
  \citenamefont {Voneshen}, \citenamefont {Kim}, \citenamefont {Jang} \emph
  {et~al.}}]{do2017incarnation}%
  \BibitemOpen
  \bibfield  {author} {\bibinfo {author} {\bibfnamefont {S.-H.}\ \bibnamefont
  {Do}}, \bibinfo {author} {\bibfnamefont {S.-Y.}\ \bibnamefont {Park}},
  \bibinfo {author} {\bibfnamefont {J.}~\bibnamefont {Yoshitake}}, \bibinfo
  {author} {\bibfnamefont {J.}~\bibnamefont {Nasu}}, \bibinfo {author}
  {\bibfnamefont {Y.}~\bibnamefont {Motome}}, \bibinfo {author} {\bibfnamefont
  {Y.~S.}\ \bibnamefont {Kwon}}, \bibinfo {author} {\bibfnamefont
  {D.}~\bibnamefont {Adroja}}, \bibinfo {author} {\bibfnamefont
  {D.}~\bibnamefont {Voneshen}}, \bibinfo {author} {\bibfnamefont
  {K.}~\bibnamefont {Kim}}, \bibinfo {author} {\bibfnamefont {T.-H.}\
  \bibnamefont {Jang}},  \emph {et~al.},\ }\href {\doibase
  10.48550/arXiv.1703.01081} {\bibfield  {journal} {\bibinfo  {journal} {arXiv
  preprint arXiv:1703.01081}\ } (\bibinfo {year} {2017}{\natexlab{a}}),\
  10.48550/arXiv.1703.01081}\BibitemShut {NoStop}%
\bibitem [{\citenamefont {Koga}\ \emph {et~al.}(2021)\citenamefont {Koga},
  \citenamefont {Murakami},\ and\ \citenamefont {Nasu}}]{koga2021majorana}%
  \BibitemOpen
  \bibfield  {author} {\bibinfo {author} {\bibfnamefont {A.}~\bibnamefont
  {Koga}}, \bibinfo {author} {\bibfnamefont {Y.}~\bibnamefont {Murakami}}, \
  and\ \bibinfo {author} {\bibfnamefont {J.}~\bibnamefont {Nasu}},\ }\href
  {\doibase 10.1103/PhysRevB.103.214421} {\bibfield  {journal} {\bibinfo
  {journal} {Physical Review B}\ }\textbf {\bibinfo {volume} {103}},\ \bibinfo
  {pages} {214421} (\bibinfo {year} {2021})}\BibitemShut {NoStop}%
\bibitem [{\citenamefont {Wulferding}\ \emph {et~al.}(2020)\citenamefont
  {Wulferding}, \citenamefont {Choi}, \citenamefont {Do}, \citenamefont {Lee},
  \citenamefont {Lemmens}, \citenamefont {Faugeras}, \citenamefont {Gallais},\
  and\ \citenamefont {Choi}}]{wulferding2020magnon}%
  \BibitemOpen
  \bibfield  {author} {\bibinfo {author} {\bibfnamefont {D.}~\bibnamefont
  {Wulferding}}, \bibinfo {author} {\bibfnamefont {Y.}~\bibnamefont {Choi}},
  \bibinfo {author} {\bibfnamefont {S.-H.}\ \bibnamefont {Do}}, \bibinfo
  {author} {\bibfnamefont {C.~H.}\ \bibnamefont {Lee}}, \bibinfo {author}
  {\bibfnamefont {P.}~\bibnamefont {Lemmens}}, \bibinfo {author} {\bibfnamefont
  {C.}~\bibnamefont {Faugeras}}, \bibinfo {author} {\bibfnamefont
  {Y.}~\bibnamefont {Gallais}}, \ and\ \bibinfo {author} {\bibfnamefont
  {K.-Y.}\ \bibnamefont {Choi}},\ }\href {\doibase 10.1038/s41467-020-15370-1}
  {\bibfield  {journal} {\bibinfo  {journal} {Nature communications}\ }\textbf
  {\bibinfo {volume} {11}},\ \bibinfo {pages} {1603} (\bibinfo {year}
  {2020})}\BibitemShut {NoStop}%
\bibitem [{\citenamefont {Hashimoto}\ \emph {et~al.}(2022)\citenamefont
  {Hashimoto}, \citenamefont {Murakami},\ and\ \citenamefont
  {Koga}}]{hashimoto2022majorana}%
  \BibitemOpen
  \bibfield  {author} {\bibinfo {author} {\bibfnamefont {A.}~\bibnamefont
  {Hashimoto}}, \bibinfo {author} {\bibfnamefont {Y.}~\bibnamefont {Murakami}},
  \ and\ \bibinfo {author} {\bibfnamefont {A.}~\bibnamefont {Koga}},\ }in\
  \href {\doibase 10.1088/1742-6596/2164/1/012028} {\emph {\bibinfo {booktitle}
  {Journal of Physics: Conference Series}}},\ Vol.\ \bibinfo {volume} {2164}\
  (\bibinfo {organization} {IOP Publishing},\ \bibinfo {year} {2022})\ p.\
  \bibinfo {pages} {012028}\BibitemShut {NoStop}%
\bibitem [{\citenamefont {Kitaev}(2006)}]{KITAEV20062}%
  \BibitemOpen
  \bibfield  {author} {\bibinfo {author} {\bibfnamefont {A.}~\bibnamefont
  {Kitaev}},\ }\href {\doibase https://doi.org/10.1016/j.aop.2005.10.005}
  {\bibfield  {journal} {\bibinfo  {journal} {Annals of Physics}\ }\textbf
  {\bibinfo {volume} {321}},\ \bibinfo {pages} {2} (\bibinfo {year} {2006})},\
  \bibinfo {note} {january Special Issue}\BibitemShut {NoStop}%
\bibitem [{\citenamefont {Hermanns}\ \emph {et~al.}(2015)\citenamefont
  {Hermanns}, \citenamefont {O’Brien},\ and\ \citenamefont
  {Trebst}}]{hermanns2015weyl}%
  \BibitemOpen
  \bibfield  {author} {\bibinfo {author} {\bibfnamefont {M.}~\bibnamefont
  {Hermanns}}, \bibinfo {author} {\bibfnamefont {K.}~\bibnamefont {O’Brien}},
  \ and\ \bibinfo {author} {\bibfnamefont {S.}~\bibnamefont {Trebst}},\ }\href
  {\doibase 10.1103/PhysRevLett.114.157202} {\bibfield  {journal} {\bibinfo
  {journal} {Physical Review Letters}\ }\textbf {\bibinfo {volume} {114}},\
  \bibinfo {pages} {157202} (\bibinfo {year} {2015})}\BibitemShut {NoStop}%
\bibitem [{\citenamefont {Le~Hur}\ \emph {et~al.}(2017)\citenamefont {Le~Hur},
  \citenamefont {Soret},\ and\ \citenamefont {Yang}}]{le2017majorana}%
  \BibitemOpen
  \bibfield  {author} {\bibinfo {author} {\bibfnamefont {K.}~\bibnamefont
  {Le~Hur}}, \bibinfo {author} {\bibfnamefont {A.}~\bibnamefont {Soret}}, \
  and\ \bibinfo {author} {\bibfnamefont {F.}~\bibnamefont {Yang}},\ }\href
  {\doibase 10.1103/PhysRevB.96.205109} {\bibfield  {journal} {\bibinfo
  {journal} {Physical Review B}\ }\textbf {\bibinfo {volume} {96}},\ \bibinfo
  {pages} {205109} (\bibinfo {year} {2017})}\BibitemShut {NoStop}%
\bibitem [{\citenamefont {Mandal}\ \emph {et~al.}(2011)\citenamefont {Mandal},
  \citenamefont {Bhattacharjee}, \citenamefont {Sengupta}, \citenamefont
  {Shankar},\ and\ \citenamefont {Baskaran}}]{mandal2011confinement}%
  \BibitemOpen
  \bibfield  {author} {\bibinfo {author} {\bibfnamefont {S.}~\bibnamefont
  {Mandal}}, \bibinfo {author} {\bibfnamefont {S.}~\bibnamefont
  {Bhattacharjee}}, \bibinfo {author} {\bibfnamefont {K.}~\bibnamefont
  {Sengupta}}, \bibinfo {author} {\bibfnamefont {R.}~\bibnamefont {Shankar}}, \
  and\ \bibinfo {author} {\bibfnamefont {G.}~\bibnamefont {Baskaran}},\ }\href
  {\doibase 10.1103/PhysRevB.84.155121} {\bibfield  {journal} {\bibinfo
  {journal} {Physical Review B}\ }\textbf {\bibinfo {volume} {84}},\ \bibinfo
  {pages} {155121} (\bibinfo {year} {2011})}\BibitemShut {NoStop}%
\bibitem [{\citenamefont {Baskaran}\ \emph {et~al.}(2007)\citenamefont
  {Baskaran}, \citenamefont {Mandal},\ and\ \citenamefont
  {Shankar}}]{2007Baskaran}%
  \BibitemOpen
  \bibfield  {author} {\bibinfo {author} {\bibfnamefont {G.}~\bibnamefont
  {Baskaran}}, \bibinfo {author} {\bibfnamefont {S.}~\bibnamefont {Mandal}}, \
  and\ \bibinfo {author} {\bibfnamefont {R.}~\bibnamefont {Shankar}},\ }\href
  {\doibase 10.1103/PhysRevLett.98.247201} {\bibfield  {journal} {\bibinfo
  {journal} {Phys. Rev. Lett.}\ }\textbf {\bibinfo {volume} {98}},\ \bibinfo
  {pages} {247201} (\bibinfo {year} {2007})}\BibitemShut {NoStop}%
\bibitem [{\citenamefont {Baskaran}\ \emph {et~al.}(2008)\citenamefont
  {Baskaran}, \citenamefont {Sen},\ and\ \citenamefont
  {Shankar}}]{2008Baskaran}%
  \BibitemOpen
  \bibfield  {author} {\bibinfo {author} {\bibfnamefont {G.}~\bibnamefont
  {Baskaran}}, \bibinfo {author} {\bibfnamefont {D.}~\bibnamefont {Sen}}, \
  and\ \bibinfo {author} {\bibfnamefont {R.}~\bibnamefont {Shankar}},\ }\href
  {\doibase 10.1103/PhysRevB.78.115116} {\bibfield  {journal} {\bibinfo
  {journal} {Phys. Rev. B}\ }\textbf {\bibinfo {volume} {78}},\ \bibinfo
  {pages} {115116} (\bibinfo {year} {2008})}\BibitemShut {NoStop}%
\bibitem [{\citenamefont {Baskaran}(2023)}]{baskaran2023metastable}%
  \BibitemOpen
  \bibfield  {author} {\bibinfo {author} {\bibfnamefont {G.}~\bibnamefont
  {Baskaran}},\ }\href@noop {} {\bibfield  {journal} {\bibinfo  {journal}
  {arXiv preprint arXiv:2309.07119}\ } (\bibinfo {year} {2023})}\BibitemShut
  {NoStop}%
\bibitem [{\citenamefont {Singh}\ \emph {et~al.}(2012)\citenamefont {Singh},
  \citenamefont {Manni}, \citenamefont {Reuther}, \citenamefont {Berlijn},
  \citenamefont {Thomale}, \citenamefont {Ku}, \citenamefont {Trebst},\ and\
  \citenamefont {Gegenwart}}]{singh2012relevance}%
  \BibitemOpen
  \bibfield  {author} {\bibinfo {author} {\bibfnamefont {Y.}~\bibnamefont
  {Singh}}, \bibinfo {author} {\bibfnamefont {S.}~\bibnamefont {Manni}},
  \bibinfo {author} {\bibfnamefont {J.}~\bibnamefont {Reuther}}, \bibinfo
  {author} {\bibfnamefont {T.}~\bibnamefont {Berlijn}}, \bibinfo {author}
  {\bibfnamefont {R.}~\bibnamefont {Thomale}}, \bibinfo {author} {\bibfnamefont
  {W.}~\bibnamefont {Ku}}, \bibinfo {author} {\bibfnamefont {S.}~\bibnamefont
  {Trebst}}, \ and\ \bibinfo {author} {\bibfnamefont {P.}~\bibnamefont
  {Gegenwart}},\ }\href {\doibase 10.1103/PhysRevLett.108.127203} {\bibfield
  {journal} {\bibinfo  {journal} {Physical review letters}\ }\textbf {\bibinfo
  {volume} {108}},\ \bibinfo {pages} {127203} (\bibinfo {year}
  {2012})}\BibitemShut {NoStop}%
\bibitem [{\citenamefont {Hwan~Chun}\ \emph {et~al.}(2015)\citenamefont
  {Hwan~Chun}, \citenamefont {Kim}, \citenamefont {Kim}, \citenamefont {Zheng},
  \citenamefont {Stoumpos}, \citenamefont {Malliakas}, \citenamefont
  {Mitchell}, \citenamefont {Mehlawat}, \citenamefont {Singh}, \citenamefont
  {Choi} \emph {et~al.}}]{hwan2015direct}%
  \BibitemOpen
  \bibfield  {author} {\bibinfo {author} {\bibfnamefont {S.}~\bibnamefont
  {Hwan~Chun}}, \bibinfo {author} {\bibfnamefont {J.-W.}\ \bibnamefont {Kim}},
  \bibinfo {author} {\bibfnamefont {J.}~\bibnamefont {Kim}}, \bibinfo {author}
  {\bibfnamefont {H.}~\bibnamefont {Zheng}}, \bibinfo {author} {\bibfnamefont
  {C.~C.}\ \bibnamefont {Stoumpos}}, \bibinfo {author} {\bibfnamefont
  {C.}~\bibnamefont {Malliakas}}, \bibinfo {author} {\bibfnamefont
  {J.}~\bibnamefont {Mitchell}}, \bibinfo {author} {\bibfnamefont
  {K.}~\bibnamefont {Mehlawat}}, \bibinfo {author} {\bibfnamefont
  {Y.}~\bibnamefont {Singh}}, \bibinfo {author} {\bibfnamefont
  {Y.}~\bibnamefont {Choi}},  \emph {et~al.},\ }\href {\doibase
  10.1038/nphys3322} {\bibfield  {journal} {\bibinfo  {journal} {Nature
  Physics}\ }\textbf {\bibinfo {volume} {11}},\ \bibinfo {pages} {462}
  (\bibinfo {year} {2015})}\BibitemShut {NoStop}%
\bibitem [{\citenamefont {Trebst}\ and\ \citenamefont
  {Hickey}(2022{\natexlab{a}})}]{TREBST20221}%
  \BibitemOpen
  \bibfield  {author} {\bibinfo {author} {\bibfnamefont {S.}~\bibnamefont
  {Trebst}}\ and\ \bibinfo {author} {\bibfnamefont {C.}~\bibnamefont
  {Hickey}},\ }\href {\doibase https://doi.org/10.1016/j.physrep.2021.11.003}
  {\bibfield  {journal} {\bibinfo  {journal} {Physics Reports}\ }\textbf
  {\bibinfo {volume} {950}},\ \bibinfo {pages} {1} (\bibinfo {year}
  {2022}{\natexlab{a}})},\ \bibinfo {note} {kitaev materials}\BibitemShut
  {NoStop}%
\bibitem [{\citenamefont {Chaloupka}\ \emph {et~al.}(2010)\citenamefont
  {Chaloupka}, \citenamefont {Jackeli},\ and\ \citenamefont
  {Khaliullin}}]{2010PRLIrridium1}%
  \BibitemOpen
  \bibfield  {author} {\bibinfo {author} {\bibfnamefont {J.~c.~v.}\
  \bibnamefont {Chaloupka}}, \bibinfo {author} {\bibfnamefont {G.}~\bibnamefont
  {Jackeli}}, \ and\ \bibinfo {author} {\bibfnamefont {G.}~\bibnamefont
  {Khaliullin}},\ }\href {\doibase 10.1103/PhysRevLett.105.027204} {\bibfield
  {journal} {\bibinfo  {journal} {Phys. Rev. Lett.}\ }\textbf {\bibinfo
  {volume} {105}},\ \bibinfo {pages} {027204} (\bibinfo {year}
  {2010})}\BibitemShut {NoStop}%
\bibitem [{\citenamefont {Suzuki}\ \emph {et~al.}(2021)\citenamefont {Suzuki},
  \citenamefont {Liu}, \citenamefont {Bertinshaw}, \citenamefont {Ueda},
  \citenamefont {Kim}, \citenamefont {Laha}, \citenamefont {Weber},
  \citenamefont {Yang}, \citenamefont {Wang}, \citenamefont {Takahashi} \emph
  {et~al.}}]{suzuki2021proximate}%
  \BibitemOpen
  \bibfield  {author} {\bibinfo {author} {\bibfnamefont {H.}~\bibnamefont
  {Suzuki}}, \bibinfo {author} {\bibfnamefont {H.}~\bibnamefont {Liu}},
  \bibinfo {author} {\bibfnamefont {J.}~\bibnamefont {Bertinshaw}}, \bibinfo
  {author} {\bibfnamefont {K.}~\bibnamefont {Ueda}}, \bibinfo {author}
  {\bibfnamefont {H.}~\bibnamefont {Kim}}, \bibinfo {author} {\bibfnamefont
  {S.}~\bibnamefont {Laha}}, \bibinfo {author} {\bibfnamefont {D.}~\bibnamefont
  {Weber}}, \bibinfo {author} {\bibfnamefont {Z.}~\bibnamefont {Yang}},
  \bibinfo {author} {\bibfnamefont {L.}~\bibnamefont {Wang}}, \bibinfo {author}
  {\bibfnamefont {H.}~\bibnamefont {Takahashi}},  \emph {et~al.},\ }\href
  {\doibase 10.1038/s41467-021-24722-4} {\bibfield  {journal} {\bibinfo
  {journal} {Nature communications}\ }\textbf {\bibinfo {volume} {12}},\
  \bibinfo {pages} {4512} (\bibinfo {year} {2021})}\BibitemShut {NoStop}%
\bibitem [{\citenamefont {Do}\ \emph {et~al.}(2017{\natexlab{b}})\citenamefont
  {Do}, \citenamefont {Park}, \citenamefont {Yoshitake}, \citenamefont {Nasu},
  \citenamefont {Motome}, \citenamefont {Kwon}, \citenamefont {Adroja},
  \citenamefont {Voneshen}, \citenamefont {Kim}, \citenamefont {Jang} \emph
  {et~al.}}]{do2017majorana}%
  \BibitemOpen
  \bibfield  {author} {\bibinfo {author} {\bibfnamefont {S.-H.}\ \bibnamefont
  {Do}}, \bibinfo {author} {\bibfnamefont {S.-Y.}\ \bibnamefont {Park}},
  \bibinfo {author} {\bibfnamefont {J.}~\bibnamefont {Yoshitake}}, \bibinfo
  {author} {\bibfnamefont {J.}~\bibnamefont {Nasu}}, \bibinfo {author}
  {\bibfnamefont {Y.}~\bibnamefont {Motome}}, \bibinfo {author} {\bibfnamefont
  {Y.~S.}\ \bibnamefont {Kwon}}, \bibinfo {author} {\bibfnamefont
  {D.}~\bibnamefont {Adroja}}, \bibinfo {author} {\bibfnamefont
  {D.}~\bibnamefont {Voneshen}}, \bibinfo {author} {\bibfnamefont
  {K.}~\bibnamefont {Kim}}, \bibinfo {author} {\bibfnamefont {T.-H.}\
  \bibnamefont {Jang}},  \emph {et~al.},\ }\href {\doibase 10.1038/nphys4264}
  {\bibfield  {journal} {\bibinfo  {journal} {Nature Physics}\ }\textbf
  {\bibinfo {volume} {13}},\ \bibinfo {pages} {1079} (\bibinfo {year}
  {2017}{\natexlab{b}})}\BibitemShut {NoStop}%
\bibitem [{\citenamefont {Banerjee}\ \emph {et~al.}(2017)\citenamefont
  {Banerjee}, \citenamefont {Yan}, \citenamefont {Knolle}, \citenamefont
  {Bridges}, \citenamefont {Stone}, \citenamefont {Lumsden}, \citenamefont
  {Mandrus}, \citenamefont {Tennant}, \citenamefont {Moessner},\ and\
  \citenamefont {Nagler}}]{banerjee2017neutron}%
  \BibitemOpen
  \bibfield  {author} {\bibinfo {author} {\bibfnamefont {A.}~\bibnamefont
  {Banerjee}}, \bibinfo {author} {\bibfnamefont {J.}~\bibnamefont {Yan}},
  \bibinfo {author} {\bibfnamefont {J.}~\bibnamefont {Knolle}}, \bibinfo
  {author} {\bibfnamefont {C.~A.}\ \bibnamefont {Bridges}}, \bibinfo {author}
  {\bibfnamefont {M.~B.}\ \bibnamefont {Stone}}, \bibinfo {author}
  {\bibfnamefont {M.~D.}\ \bibnamefont {Lumsden}}, \bibinfo {author}
  {\bibfnamefont {D.~G.}\ \bibnamefont {Mandrus}}, \bibinfo {author}
  {\bibfnamefont {D.~A.}\ \bibnamefont {Tennant}}, \bibinfo {author}
  {\bibfnamefont {R.}~\bibnamefont {Moessner}}, \ and\ \bibinfo {author}
  {\bibfnamefont {S.~E.}\ \bibnamefont {Nagler}},\ }\href {\doibase
  10.1126/science.aah6015} {\bibfield  {journal} {\bibinfo  {journal}
  {Science}\ }\textbf {\bibinfo {volume} {356}},\ \bibinfo {pages} {1055}
  (\bibinfo {year} {2017})}\BibitemShut {NoStop}%
\bibitem [{\citenamefont {Trebst}\ and\ \citenamefont
  {Hickey}(2022{\natexlab{b}})}]{trebst2022kitaev}%
  \BibitemOpen
  \bibfield  {author} {\bibinfo {author} {\bibfnamefont {S.}~\bibnamefont
  {Trebst}}\ and\ \bibinfo {author} {\bibfnamefont {C.}~\bibnamefont
  {Hickey}},\ }\href {\doibase 10.1016/j.physrep.2021.11.003} {\bibfield
  {journal} {\bibinfo  {journal} {Physics Reports}\ }\textbf {\bibinfo {volume}
  {950}},\ \bibinfo {pages} {1} (\bibinfo {year}
  {2022}{\natexlab{b}})}\BibitemShut {NoStop}%
\bibitem [{\citenamefont {Winter}\ \emph {et~al.}(2016)\citenamefont {Winter},
  \citenamefont {Li}, \citenamefont {Jeschke},\ and\ \citenamefont
  {Valent{\'\i}}}]{winter2016challenges}%
  \BibitemOpen
  \bibfield  {author} {\bibinfo {author} {\bibfnamefont {S.~M.}\ \bibnamefont
  {Winter}}, \bibinfo {author} {\bibfnamefont {Y.}~\bibnamefont {Li}}, \bibinfo
  {author} {\bibfnamefont {H.~O.}\ \bibnamefont {Jeschke}}, \ and\ \bibinfo
  {author} {\bibfnamefont {R.}~\bibnamefont {Valent{\'\i}}},\ }\href {\doibase
  10.1103/PhysRevB.93.214431} {\bibfield  {journal} {\bibinfo  {journal}
  {Physical Review B}\ }\textbf {\bibinfo {volume} {93}},\ \bibinfo {pages}
  {214431} (\bibinfo {year} {2016})}\BibitemShut {NoStop}%
\bibitem [{\citenamefont {Janssen}\ \emph {et~al.}(2016)\citenamefont
  {Janssen}, \citenamefont {Andrade},\ and\ \citenamefont
  {Vojta}}]{janssen2016honeycomb}%
  \BibitemOpen
  \bibfield  {author} {\bibinfo {author} {\bibfnamefont {L.}~\bibnamefont
  {Janssen}}, \bibinfo {author} {\bibfnamefont {E.~C.}\ \bibnamefont
  {Andrade}}, \ and\ \bibinfo {author} {\bibfnamefont {M.}~\bibnamefont
  {Vojta}},\ }\href {\doibase 10.1103/PhysRevLett.117.277202} {\bibfield
  {journal} {\bibinfo  {journal} {Phys. Rev. Lett.}\ }\textbf {\bibinfo
  {volume} {117}},\ \bibinfo {pages} {277202} (\bibinfo {year}
  {2016})}\BibitemShut {NoStop}%
\bibitem [{\citenamefont {Katukuri}\ \emph {et~al.}(2016)\citenamefont
  {Katukuri}, \citenamefont {Yadav}, \citenamefont {Hozoi}, \citenamefont
  {Nishimoto},\ and\ \citenamefont {van~den Brink}}]{katukuri2016vicinity}%
  \BibitemOpen
  \bibfield  {author} {\bibinfo {author} {\bibfnamefont {V.~M.}\ \bibnamefont
  {Katukuri}}, \bibinfo {author} {\bibfnamefont {R.}~\bibnamefont {Yadav}},
  \bibinfo {author} {\bibfnamefont {L.}~\bibnamefont {Hozoi}}, \bibinfo
  {author} {\bibfnamefont {S.}~\bibnamefont {Nishimoto}}, \ and\ \bibinfo
  {author} {\bibfnamefont {J.}~\bibnamefont {van~den Brink}},\ }\href {\doibase
  10.1038/srep29585} {\bibfield  {journal} {\bibinfo  {journal} {Scientific
  reports}\ }\textbf {\bibinfo {volume} {6}},\ \bibinfo {pages} {29585}
  (\bibinfo {year} {2016})}\BibitemShut {NoStop}%
\bibitem [{\citenamefont {Takayama}\ \emph {et~al.}(2015)\citenamefont
  {Takayama}, \citenamefont {Kato}, \citenamefont {Dinnebier}, \citenamefont
  {Nuss}, \citenamefont {Kono}, \citenamefont {Veiga}, \citenamefont {Fabbris},
  \citenamefont {Haskel},\ and\ \citenamefont
  {Takagi}}]{takayama2015hyperhoneycomb}%
  \BibitemOpen
  \bibfield  {author} {\bibinfo {author} {\bibfnamefont {T.}~\bibnamefont
  {Takayama}}, \bibinfo {author} {\bibfnamefont {A.}~\bibnamefont {Kato}},
  \bibinfo {author} {\bibfnamefont {R.}~\bibnamefont {Dinnebier}}, \bibinfo
  {author} {\bibfnamefont {J.}~\bibnamefont {Nuss}}, \bibinfo {author}
  {\bibfnamefont {H.}~\bibnamefont {Kono}}, \bibinfo {author} {\bibfnamefont
  {L.}~\bibnamefont {Veiga}}, \bibinfo {author} {\bibfnamefont
  {G.}~\bibnamefont {Fabbris}}, \bibinfo {author} {\bibfnamefont
  {D.}~\bibnamefont {Haskel}}, \ and\ \bibinfo {author} {\bibfnamefont
  {H.}~\bibnamefont {Takagi}},\ }\href {\doibase
  10.1103/PhysRevLett.114.077202} {\bibfield  {journal} {\bibinfo  {journal}
  {Physical review letters}\ }\textbf {\bibinfo {volume} {114}},\ \bibinfo
  {pages} {077202} (\bibinfo {year} {2015})}\BibitemShut {NoStop}%
\bibitem [{\citenamefont {Tsirlin}\ and\ \citenamefont
  {Gegenwart}(2022)}]{tsirlin2022kitaev}%
  \BibitemOpen
  \bibfield  {author} {\bibinfo {author} {\bibfnamefont {A.~A.}\ \bibnamefont
  {Tsirlin}}\ and\ \bibinfo {author} {\bibfnamefont {P.}~\bibnamefont
  {Gegenwart}},\ }\href {\doibase 10.1002/pssb.202100146} {\bibfield  {journal}
  {\bibinfo  {journal} {physica status solidi (b)}\ }\textbf {\bibinfo {volume}
  {259}},\ \bibinfo {pages} {2100146} (\bibinfo {year} {2022})}\BibitemShut
  {NoStop}%
\bibitem [{\citenamefont {Gotfryd}\ \emph {et~al.}(2017)\citenamefont
  {Gotfryd}, \citenamefont {Rusna{\v{c}}ko}, \citenamefont {Wohlfeld},
  \citenamefont {Jackeli}, \citenamefont {Chaloupka},\ and\ \citenamefont
  {Ole{\'s}}}]{gotfryd2017phase}%
  \BibitemOpen
  \bibfield  {author} {\bibinfo {author} {\bibfnamefont {D.}~\bibnamefont
  {Gotfryd}}, \bibinfo {author} {\bibfnamefont {J.}~\bibnamefont
  {Rusna{\v{c}}ko}}, \bibinfo {author} {\bibfnamefont {K.}~\bibnamefont
  {Wohlfeld}}, \bibinfo {author} {\bibfnamefont {G.}~\bibnamefont {Jackeli}},
  \bibinfo {author} {\bibfnamefont {J.}~\bibnamefont {Chaloupka}}, \ and\
  \bibinfo {author} {\bibfnamefont {A.~M.}\ \bibnamefont {Ole{\'s}}},\ }\href
  {\doibase 10.1103/PhysRevB.95.024426} {\bibfield  {journal} {\bibinfo
  {journal} {Physical Review B}\ }\textbf {\bibinfo {volume} {95}},\ \bibinfo
  {pages} {024426} (\bibinfo {year} {2017})}\BibitemShut {NoStop}%
\bibitem [{\citenamefont {Schaffer}\ \emph {et~al.}(2012)\citenamefont
  {Schaffer}, \citenamefont {Bhattacharjee},\ and\ \citenamefont
  {Kim}}]{schaffer2012quantum}%
  \BibitemOpen
  \bibfield  {author} {\bibinfo {author} {\bibfnamefont {R.}~\bibnamefont
  {Schaffer}}, \bibinfo {author} {\bibfnamefont {S.}~\bibnamefont
  {Bhattacharjee}}, \ and\ \bibinfo {author} {\bibfnamefont {Y.~B.}\
  \bibnamefont {Kim}},\ }\href {\doibase 10.1103/PhysRevB.86.224417} {\bibfield
   {journal} {\bibinfo  {journal} {Physical Review B}\ }\textbf {\bibinfo
  {volume} {86}},\ \bibinfo {pages} {224417} (\bibinfo {year}
  {2012})}\BibitemShut {NoStop}%
\bibitem [{\citenamefont {Feng}\ \emph {et~al.}(2007)\citenamefont {Feng},
  \citenamefont {Zhang},\ and\ \citenamefont {Xiang}}]{feng2007topological}%
  \BibitemOpen
  \bibfield  {author} {\bibinfo {author} {\bibfnamefont {X.-Y.}\ \bibnamefont
  {Feng}}, \bibinfo {author} {\bibfnamefont {G.-M.}\ \bibnamefont {Zhang}}, \
  and\ \bibinfo {author} {\bibfnamefont {T.}~\bibnamefont {Xiang}},\ }\href
  {\doibase 10.1103/PhysRevLett.98.087204} {\bibfield  {journal} {\bibinfo
  {journal} {Phys. Rev. Lett.}\ }\textbf {\bibinfo {volume} {98}},\ \bibinfo
  {pages} {087204} (\bibinfo {year} {2007})}\BibitemShut {NoStop}%
\bibitem [{\citenamefont {Holleis}\ \emph {et~al.}(2021)\citenamefont
  {Holleis}, \citenamefont {Prestigiacomo}, \citenamefont {Fan}, \citenamefont
  {Nishimoto}, \citenamefont {Osofsky}, \citenamefont {Chern}, \citenamefont
  {van~den Brink},\ and\ \citenamefont {Shivaram}}]{holleis2021anomalous}%
  \BibitemOpen
  \bibfield  {author} {\bibinfo {author} {\bibfnamefont {L.}~\bibnamefont
  {Holleis}}, \bibinfo {author} {\bibfnamefont {J.~C.}\ \bibnamefont
  {Prestigiacomo}}, \bibinfo {author} {\bibfnamefont {Z.}~\bibnamefont {Fan}},
  \bibinfo {author} {\bibfnamefont {S.}~\bibnamefont {Nishimoto}}, \bibinfo
  {author} {\bibfnamefont {M.}~\bibnamefont {Osofsky}}, \bibinfo {author}
  {\bibfnamefont {G.-W.}\ \bibnamefont {Chern}}, \bibinfo {author}
  {\bibfnamefont {J.}~\bibnamefont {van~den Brink}}, \ and\ \bibinfo {author}
  {\bibfnamefont {B.}~\bibnamefont {Shivaram}},\ }\href@noop {} {\bibfield
  {journal} {\bibinfo  {journal} {npj Quantum Materials}\ }\textbf {\bibinfo
  {volume} {6}},\ \bibinfo {pages} {66} (\bibinfo {year} {2021})}\BibitemShut
  {NoStop}%
\bibitem [{\citenamefont {Kelker}(1973)}]{1973kelker}%
  \BibitemOpen
  \bibfield  {author} {\bibinfo {author} {\bibfnamefont {H.}~\bibnamefont
  {Kelker}},\ }\href {\doibase 10.1080/15421407308083312} {\bibfield  {journal}
  {\bibinfo  {journal} {Molecular Crystals and Liquid Crystals}\ }\textbf
  {\bibinfo {volume} {21}},\ \bibinfo {pages} {1} (\bibinfo {year} {1973})},\
  \Eprint {http://arxiv.org/abs/https://doi.org/10.1080/15421407308083312}
  {https://doi.org/10.1080/15421407308083312} \BibitemShut {NoStop}%
\bibitem [{\citenamefont {Kelker}(1988)}]{1988kelker}%
  \BibitemOpen
  \bibfield  {author} {\bibinfo {author} {\bibfnamefont {H.}~\bibnamefont
  {Kelker}},\ }\href {\doibase 10.1080/00268948808082195} {\bibfield  {journal}
  {\bibinfo  {journal} {Molecular Crystals and Liquid Crystals Incorporating
  Nonlinear Optics}\ }\textbf {\bibinfo {volume} {165}},\ \bibinfo {pages} {1}
  (\bibinfo {year} {1988})},\ \Eprint
  {http://arxiv.org/abs/https://doi.org/10.1080/00268948808082195}
  {https://doi.org/10.1080/00268948808082195} \BibitemShut {NoStop}%
\bibitem [{\citenamefont {Mermin}\ and\ \citenamefont
  {Wagner}(1966)}]{1966MerminWagner}%
  \BibitemOpen
  \bibfield  {author} {\bibinfo {author} {\bibfnamefont {N.~D.}\ \bibnamefont
  {Mermin}}\ and\ \bibinfo {author} {\bibfnamefont {H.}~\bibnamefont
  {Wagner}},\ }\href {\doibase 10.1103/PhysRevLett.17.1133} {\bibfield
  {journal} {\bibinfo  {journal} {Phys. Rev. Lett.}\ }\textbf {\bibinfo
  {volume} {17}},\ \bibinfo {pages} {1133} (\bibinfo {year}
  {1966})}\BibitemShut {NoStop}%
\bibitem [{\citenamefont {Alexander}\ \emph {et~al.}(2012)\citenamefont
  {Alexander}, \citenamefont {Chen}, \citenamefont {Matsumoto},\ and\
  \citenamefont {Kamien}}]{2012liquidcrystalRevModPhys}%
  \BibitemOpen
  \bibfield  {author} {\bibinfo {author} {\bibfnamefont {G.~P.}\ \bibnamefont
  {Alexander}}, \bibinfo {author} {\bibfnamefont {B.~G.-g.}\ \bibnamefont
  {Chen}}, \bibinfo {author} {\bibfnamefont {E.~A.}\ \bibnamefont {Matsumoto}},
  \ and\ \bibinfo {author} {\bibfnamefont {R.~D.}\ \bibnamefont {Kamien}},\
  }\href {\doibase 10.1103/RevModPhys.84.497} {\bibfield  {journal} {\bibinfo
  {journal} {Rev. Mod. Phys.}\ }\textbf {\bibinfo {volume} {84}},\ \bibinfo
  {pages} {497} (\bibinfo {year} {2012})}\BibitemShut {NoStop}%
\bibitem [{\citenamefont {L\"auchli}\ \emph {et~al.}(2006)\citenamefont
  {L\"auchli}, \citenamefont {Mila},\ and\ \citenamefont
  {Penc}}]{2006lauchili}%
  \BibitemOpen
  \bibfield  {author} {\bibinfo {author} {\bibfnamefont {A.}~\bibnamefont
  {L\"auchli}}, \bibinfo {author} {\bibfnamefont {F.}~\bibnamefont {Mila}}, \
  and\ \bibinfo {author} {\bibfnamefont {K.}~\bibnamefont {Penc}},\ }\href
  {\doibase 10.1103/PhysRevLett.97.087205} {\bibfield  {journal} {\bibinfo
  {journal} {Phys. Rev. Lett.}\ }\textbf {\bibinfo {volume} {97}},\ \bibinfo
  {pages} {087205} (\bibinfo {year} {2006})}\BibitemShut {NoStop}%
\bibitem [{\citenamefont {Tsunetsugu}\ and\ \citenamefont
  {Arikawa}(2006)}]{2006Tsunetsugu}%
  \BibitemOpen
  \bibfield  {author} {\bibinfo {author} {\bibfnamefont {H.}~\bibnamefont
  {Tsunetsugu}}\ and\ \bibinfo {author} {\bibfnamefont {M.}~\bibnamefont
  {Arikawa}},\ }\href {\doibase 10.1143/JPSJ.75.083701} {\bibfield  {journal}
  {\bibinfo  {journal} {Journal of the Physical Society of Japan}\ }\textbf
  {\bibinfo {volume} {75}},\ \bibinfo {pages} {083701} (\bibinfo {year}
  {2006})},\ \Eprint
  {http://arxiv.org/abs/https://doi.org/10.1143/JPSJ.75.083701}
  {https://doi.org/10.1143/JPSJ.75.083701} \BibitemShut {NoStop}%
\bibitem [{\citenamefont {Lacroix}\ \emph {et~al.}(2011)\citenamefont
  {Lacroix}, \citenamefont {Mendels},\ and\ \citenamefont
  {Mila}}]{lacroix2011introduction}%
  \BibitemOpen
  \bibfield  {author} {\bibinfo {author} {\bibfnamefont {C.}~\bibnamefont
  {Lacroix}}, \bibinfo {author} {\bibfnamefont {P.}~\bibnamefont {Mendels}}, \
  and\ \bibinfo {author} {\bibfnamefont {F.}~\bibnamefont {Mila}},\ }\href@noop
  {} {\emph {\bibinfo {title} {Introduction to frustrated magnetism: materials,
  experiments, theory}}},\ Vol.\ \bibinfo {volume} {164}\ (\bibinfo
  {publisher} {Springer Science \& Business Media},\ \bibinfo {year}
  {2011})\BibitemShut {NoStop}%
\bibitem [{\citenamefont {Blume}\ and\ \citenamefont
  {Hsieh}(1969)}]{blume1969biquadratic}%
  \BibitemOpen
  \bibfield  {author} {\bibinfo {author} {\bibfnamefont {M.}~\bibnamefont
  {Blume}}\ and\ \bibinfo {author} {\bibfnamefont {Y.}~\bibnamefont {Hsieh}},\
  }\href {https://doi.org/10.1063/1.1657616} {\bibfield  {journal} {\bibinfo
  {journal} {Journal of Applied Physics}\ }\textbf {\bibinfo {volume} {40}},\
  \bibinfo {pages} {1249} (\bibinfo {year} {1969})}\BibitemShut {NoStop}%
\bibitem [{\citenamefont {Hu}\ \emph {et~al.}(2019)\citenamefont {Hu},
  \citenamefont {Gong}, \citenamefont {Lai}, \citenamefont {Hu}, \citenamefont
  {Si},\ and\ \citenamefont {Nevidomskyy}}]{hu2019nematic}%
  \BibitemOpen
  \bibfield  {author} {\bibinfo {author} {\bibfnamefont {W.-J.}\ \bibnamefont
  {Hu}}, \bibinfo {author} {\bibfnamefont {S.-S.}\ \bibnamefont {Gong}},
  \bibinfo {author} {\bibfnamefont {H.-H.}\ \bibnamefont {Lai}}, \bibinfo
  {author} {\bibfnamefont {H.}~\bibnamefont {Hu}}, \bibinfo {author}
  {\bibfnamefont {Q.}~\bibnamefont {Si}}, \ and\ \bibinfo {author}
  {\bibfnamefont {A.~H.}\ \bibnamefont {Nevidomskyy}},\ }\href {\doibase
  10.1103/PhysRevB.100.165142} {\bibfield  {journal} {\bibinfo  {journal}
  {Phys. Rev. B}\ }\textbf {\bibinfo {volume} {100}},\ \bibinfo {pages}
  {165142} (\bibinfo {year} {2019})}\BibitemShut {NoStop}%
\bibitem [{\citenamefont {Podolsky}\ and\ \citenamefont
  {Demler}(2005)}]{podolsky2005properties}%
  \BibitemOpen
  \bibfield  {author} {\bibinfo {author} {\bibfnamefont {D.}~\bibnamefont
  {Podolsky}}\ and\ \bibinfo {author} {\bibfnamefont {E.}~\bibnamefont
  {Demler}},\ }\href {\doibase 10.1088/1367-2630/7/1/059} {\bibfield  {journal}
  {\bibinfo  {journal} {New Journal of Physics}\ }\textbf {\bibinfo {volume}
  {7}},\ \bibinfo {pages} {59} (\bibinfo {year} {2005})}\BibitemShut {NoStop}%
\bibitem [{\citenamefont {Tanaka}\ and\ \citenamefont
  {Hotta}(2020)}]{tanaka2020multiple}%
  \BibitemOpen
  \bibfield  {author} {\bibinfo {author} {\bibfnamefont {K.}~\bibnamefont
  {Tanaka}}\ and\ \bibinfo {author} {\bibfnamefont {C.}~\bibnamefont {Hotta}},\
  }\href {\doibase 10.1103/PhysRevB.101.094422} {\bibfield  {journal} {\bibinfo
   {journal} {Phys. Rev. B}\ }\textbf {\bibinfo {volume} {101}},\ \bibinfo
  {pages} {094422} (\bibinfo {year} {2020})}\BibitemShut {NoStop}%
\bibitem [{\citenamefont {Hikihara}\ \emph {et~al.}(2008)\citenamefont
  {Hikihara}, \citenamefont {Kecke}, \citenamefont {Momoi},\ and\ \citenamefont
  {Furusaki}}]{hikihara2008vector}%
  \BibitemOpen
  \bibfield  {author} {\bibinfo {author} {\bibfnamefont {T.}~\bibnamefont
  {Hikihara}}, \bibinfo {author} {\bibfnamefont {L.}~\bibnamefont {Kecke}},
  \bibinfo {author} {\bibfnamefont {T.}~\bibnamefont {Momoi}}, \ and\ \bibinfo
  {author} {\bibfnamefont {A.}~\bibnamefont {Furusaki}},\ }\href {\doibase
  10.1103/PhysRevB.78.144404} {\bibfield  {journal} {\bibinfo  {journal} {Phys.
  Rev. B}\ }\textbf {\bibinfo {volume} {78}},\ \bibinfo {pages} {144404}
  (\bibinfo {year} {2008})}\BibitemShut {NoStop}%
\bibitem [{\citenamefont {Parvej}\ and\ \citenamefont
  {Kumar}(2017)}]{parvej2017multipolar}%
  \BibitemOpen
  \bibfield  {author} {\bibinfo {author} {\bibfnamefont {A.}~\bibnamefont
  {Parvej}}\ and\ \bibinfo {author} {\bibfnamefont {M.}~\bibnamefont {Kumar}},\
  }\href {\doibase 10.1103/PhysRevB.96.054413} {\bibfield  {journal} {\bibinfo
  {journal} {Phys. Rev. B}\ }\textbf {\bibinfo {volume} {96}},\ \bibinfo
  {pages} {054413} (\bibinfo {year} {2017})}\BibitemShut {NoStop}%
\bibitem [{\citenamefont {Sudan}\ \emph {et~al.}(2009)\citenamefont {Sudan},
  \citenamefont {L\"uscher},\ and\ \citenamefont
  {L\"auchli}}]{sudan2009emergent}%
  \BibitemOpen
  \bibfield  {author} {\bibinfo {author} {\bibfnamefont {J.}~\bibnamefont
  {Sudan}}, \bibinfo {author} {\bibfnamefont {A.}~\bibnamefont {L\"uscher}}, \
  and\ \bibinfo {author} {\bibfnamefont {A.~M.}\ \bibnamefont {L\"auchli}},\
  }\href {\doibase 10.1103/PhysRevB.80.140402} {\bibfield  {journal} {\bibinfo
  {journal} {Phys. Rev. B}\ }\textbf {\bibinfo {volume} {80}},\ \bibinfo
  {pages} {140402} (\bibinfo {year} {2009})}\BibitemShut {NoStop}%
\bibitem [{\citenamefont {Nasu}\ \emph {et~al.}(2017)\citenamefont {Nasu},
  \citenamefont {Kato}, \citenamefont {Yoshitake}, \citenamefont {Kamiya},\
  and\ \citenamefont {Motome}}]{2017PRLNasu}%
  \BibitemOpen
  \bibfield  {author} {\bibinfo {author} {\bibfnamefont {J.}~\bibnamefont
  {Nasu}}, \bibinfo {author} {\bibfnamefont {Y.}~\bibnamefont {Kato}}, \bibinfo
  {author} {\bibfnamefont {J.}~\bibnamefont {Yoshitake}}, \bibinfo {author}
  {\bibfnamefont {Y.}~\bibnamefont {Kamiya}}, \ and\ \bibinfo {author}
  {\bibfnamefont {Y.}~\bibnamefont {Motome}},\ }\href {\doibase
  10.1103/PhysRevLett.118.137203} {\bibfield  {journal} {\bibinfo  {journal}
  {Phys. Rev. Lett.}\ }\textbf {\bibinfo {volume} {118}},\ \bibinfo {pages}
  {137203} (\bibinfo {year} {2017})}\BibitemShut {NoStop}%
\bibitem [{\citenamefont {Takahashi}\ \emph {et~al.}(2021)\citenamefont
  {Takahashi}, \citenamefont {Yamada}, \citenamefont {Takikawa}, \citenamefont
  {Mizushima},\ and\ \citenamefont {Fujimoto}}]{2021PRRTakahashi}%
  \BibitemOpen
  \bibfield  {author} {\bibinfo {author} {\bibfnamefont {M.~O.}\ \bibnamefont
  {Takahashi}}, \bibinfo {author} {\bibfnamefont {M.~G.}\ \bibnamefont
  {Yamada}}, \bibinfo {author} {\bibfnamefont {D.}~\bibnamefont {Takikawa}},
  \bibinfo {author} {\bibfnamefont {T.}~\bibnamefont {Mizushima}}, \ and\
  \bibinfo {author} {\bibfnamefont {S.}~\bibnamefont {Fujimoto}},\ }\href
  {\doibase 10.1103/PhysRevResearch.3.023189} {\bibfield  {journal} {\bibinfo
  {journal} {Phys. Rev. Res.}\ }\textbf {\bibinfo {volume} {3}},\ \bibinfo
  {pages} {023189} (\bibinfo {year} {2021})}\BibitemShut {NoStop}%
\bibitem [{\citenamefont {Penc}\ and\ \citenamefont
  {L{\"a}uchli}(2011)}]{Penc2011}%
  \BibitemOpen
  \bibfield  {author} {\bibinfo {author} {\bibfnamefont {K.}~\bibnamefont
  {Penc}}\ and\ \bibinfo {author} {\bibfnamefont {A.~M.}\ \bibnamefont
  {L{\"a}uchli}},\ }\enquote {\bibinfo {title} {Spin nematic phases in quantum
  spin systems},}\ in\ \href {\doibase 10.1007/978-3-642-10589-0_13} {\emph
  {\bibinfo {booktitle} {Introduction to Frustrated Magnetism: Materials,
  Experiments, Theory}}},\ \bibinfo {editor} {edited by\ \bibinfo {editor}
  {\bibfnamefont {C.}~\bibnamefont {Lacroix}}, \bibinfo {editor} {\bibfnamefont
  {P.}~\bibnamefont {Mendels}}, \ and\ \bibinfo {editor} {\bibfnamefont
  {F.}~\bibnamefont {Mila}}}\ (\bibinfo  {publisher} {Springer Berlin
  Heidelberg},\ \bibinfo {address} {Berlin, Heidelberg},\ \bibinfo {year}
  {2011})\ pp.\ \bibinfo {pages} {331--362}\BibitemShut {NoStop}%
\bibitem [{\citenamefont {Grafe}\ \emph {et~al.}(2017)\citenamefont {Grafe},
  \citenamefont {Nishimoto}, \citenamefont {Iakovleva}, \citenamefont
  {Vavilova}, \citenamefont {Spillecke}, \citenamefont {Alfonsov},
  \citenamefont {Sturza}, \citenamefont {Wurmehl}, \citenamefont {Nojiri},
  \citenamefont {Rosner} \emph {et~al.}}]{grafe2017signatures}%
  \BibitemOpen
  \bibfield  {author} {\bibinfo {author} {\bibfnamefont {H.-J.}\ \bibnamefont
  {Grafe}}, \bibinfo {author} {\bibfnamefont {S.}~\bibnamefont {Nishimoto}},
  \bibinfo {author} {\bibfnamefont {M.}~\bibnamefont {Iakovleva}}, \bibinfo
  {author} {\bibfnamefont {E.}~\bibnamefont {Vavilova}}, \bibinfo {author}
  {\bibfnamefont {L.}~\bibnamefont {Spillecke}}, \bibinfo {author}
  {\bibfnamefont {A.}~\bibnamefont {Alfonsov}}, \bibinfo {author}
  {\bibfnamefont {M.-I.}\ \bibnamefont {Sturza}}, \bibinfo {author}
  {\bibfnamefont {S.}~\bibnamefont {Wurmehl}}, \bibinfo {author} {\bibfnamefont
  {H.}~\bibnamefont {Nojiri}}, \bibinfo {author} {\bibfnamefont
  {H.}~\bibnamefont {Rosner}},  \emph {et~al.},\ }\href {\doibase
  10.1038/s41598-017-06525-0} {\bibfield  {journal} {\bibinfo  {journal}
  {Scientific reports}\ }\textbf {\bibinfo {volume} {7}},\ \bibinfo {pages}
  {6720} (\bibinfo {year} {2017})}\BibitemShut {NoStop}%
\bibitem [{\citenamefont {Agrapidis}\ \emph {et~al.}(2019)\citenamefont
  {Agrapidis}, \citenamefont {van~den Brink},\ and\ \citenamefont
  {Nishimoto}}]{agrapidis2019ground}%
  \BibitemOpen
  \bibfield  {author} {\bibinfo {author} {\bibfnamefont {C.~E.}\ \bibnamefont
  {Agrapidis}}, \bibinfo {author} {\bibfnamefont {J.}~\bibnamefont {van~den
  Brink}}, \ and\ \bibinfo {author} {\bibfnamefont {S.}~\bibnamefont
  {Nishimoto}},\ }\href {\doibase 10.1103/PhysRevB.99.224418} {\bibfield
  {journal} {\bibinfo  {journal} {Phys. Rev. B}\ }\textbf {\bibinfo {volume}
  {99}},\ \bibinfo {pages} {224418} (\bibinfo {year} {2019})}\BibitemShut
  {NoStop}%
\bibitem [{\citenamefont {Catuneanu}\ \emph {et~al.}(2019)\citenamefont
  {Catuneanu}, \citenamefont {S\o{}rensen},\ and\ \citenamefont
  {Kee}}]{catuneanu2019nonlocal}%
  \BibitemOpen
  \bibfield  {author} {\bibinfo {author} {\bibfnamefont {A.}~\bibnamefont
  {Catuneanu}}, \bibinfo {author} {\bibfnamefont {E.~S.}\ \bibnamefont
  {S\o{}rensen}}, \ and\ \bibinfo {author} {\bibfnamefont {H.-Y.}\ \bibnamefont
  {Kee}},\ }\href {\doibase 10.1103/PhysRevB.99.195112} {\bibfield  {journal}
  {\bibinfo  {journal} {Phys. Rev. B}\ }\textbf {\bibinfo {volume} {99}},\
  \bibinfo {pages} {195112} (\bibinfo {year} {2019})}\BibitemShut {NoStop}%
\bibitem [{sup()}]{suppmat}%
  \BibitemOpen
  \href@noop {} {}\bibinfo {note} {Supplemental materials on the following
  topics in relation to the main text: (I) Schematic picture of quadrupolar
  ordered states, (II) Wave function analysis in two-leg ladder, (III)
  Quadrupolar order and correlation in 2-leg ladder from ED, (IV) Finite-size
  effects in the quadrupolar correlation functions in 2-leg
  ladder.}\BibitemShut {Stop}%
\bibitem [{\citenamefont {Shannon}\ \emph {et~al.}(2006)\citenamefont
  {Shannon}, \citenamefont {Momoi},\ and\ \citenamefont
  {Sindzingre}}]{Shannon2006}%
  \BibitemOpen
  \bibfield  {author} {\bibinfo {author} {\bibfnamefont {N.}~\bibnamefont
  {Shannon}}, \bibinfo {author} {\bibfnamefont {T.}~\bibnamefont {Momoi}}, \
  and\ \bibinfo {author} {\bibfnamefont {P.}~\bibnamefont {Sindzingre}},\
  }\href {\doibase 10.1103/PhysRevLett.96.027213} {\bibfield  {journal}
  {\bibinfo  {journal} {Phys. Rev. Lett.}\ }\textbf {\bibinfo {volume} {96}},\
  \bibinfo {pages} {027213} (\bibinfo {year} {2006})}\BibitemShut {NoStop}%
\bibitem [{\citenamefont {Yadav}\ \emph {et~al.}(2016)\citenamefont {Yadav},
  \citenamefont {Bogdanov}, \citenamefont {Katukuri}, \citenamefont
  {Nishimoto}, \citenamefont {Van Den~Brink},\ and\ \citenamefont
  {Hozoi}}]{yadav2016kitaev}%
  \BibitemOpen
  \bibfield  {author} {\bibinfo {author} {\bibfnamefont {R.}~\bibnamefont
  {Yadav}}, \bibinfo {author} {\bibfnamefont {N.~A.}\ \bibnamefont {Bogdanov}},
  \bibinfo {author} {\bibfnamefont {V.~M.}\ \bibnamefont {Katukuri}}, \bibinfo
  {author} {\bibfnamefont {S.}~\bibnamefont {Nishimoto}}, \bibinfo {author}
  {\bibfnamefont {J.}~\bibnamefont {Van Den~Brink}}, \ and\ \bibinfo {author}
  {\bibfnamefont {L.}~\bibnamefont {Hozoi}},\ }\href@noop {} {\bibfield
  {journal} {\bibinfo  {journal} {Scientific reports}\ }\textbf {\bibinfo
  {volume} {6}},\ \bibinfo {pages} {37925} (\bibinfo {year}
  {2016})}\BibitemShut {NoStop}%
\end{thebibliography}%


\begin{thebibliography}{3}%
\makeatletter
\providecommand \@ifxundefined [1]{%
 \@ifx{#1\undefined}
}%
\providecommand \@ifnum [1]{%
 \ifnum #1\expandafter \@firstoftwo
 \else \expandafter \@secondoftwo
 \fi
}%
\providecommand \@ifx [1]{%
 \ifx #1\expandafter \@firstoftwo
 \else \expandafter \@secondoftwo
 \fi
}%
\providecommand \natexlab [1]{#1}%
\providecommand \enquote  [1]{``#1''}%
\providecommand \bibnamefont  [1]{#1}%
\providecommand \bibfnamefont [1]{#1}%
\providecommand \citenamefont [1]{#1}%
\providecommand \href@noop [0]{\@secondoftwo}%
\providecommand \href [0]{\begingroup \@sanitize@url \@href}%
\providecommand \@href[1]{\@@startlink{#1}\@@href}%
\providecommand \@@href[1]{\endgroup#1\@@endlink}%
\providecommand \@sanitize@url [0]{\catcode `\\12\catcode `\$12\catcode
  `\&12\catcode `\#12\catcode `\^12\catcode `\_12\catcode `\%12\relax}%
\providecommand \@@startlink[1]{}%
\providecommand \@@endlink[0]{}%
\providecommand \url  [0]{\begingroup\@sanitize@url \@url }%
\providecommand \@url [1]{\endgroup\@href {#1}{\urlprefix }}%
\providecommand \urlprefix  [0]{URL }%
\providecommand \Eprint [0]{\href }%
\providecommand \doibase [0]{http://dx.doi.org/}%
\providecommand \selectlanguage [0]{\@gobble}%
\providecommand \bibinfo  [0]{\@secondoftwo}%
\providecommand \bibfield  [0]{\@secondoftwo}%
\providecommand \translation [1]{[#1]}%
\providecommand \BibitemOpen [0]{}%
\providecommand \bibitemStop [0]{}%
\providecommand \bibitemNoStop [0]{.\EOS\space}%
\providecommand \EOS [0]{\spacefactor3000\relax}%
\providecommand \BibitemShut  [1]{\csname bibitem#1\endcsname}%
\let\auto@bib@innerbib\@empty
\bibitem [{\citenamefont {White}(1992)}]{white1992}%
  \BibitemOpen
  \bibfield  {author} {\bibinfo {author} {\bibfnamefont {S.~R.}\ \bibnamefont
  {White}},\ }\href {\doibase 10.1103/PhysRevLett.69.2863} {\bibfield
  {journal} {\bibinfo  {journal} {Phys. Rev. Lett.}\ }\textbf {\bibinfo
  {volume} {69}},\ \bibinfo {pages} {2863} (\bibinfo {year}
  {1992})}\BibitemShut {NoStop}%
\bibitem [{\citenamefont {White}(1993)}]{white1993}%
  \BibitemOpen
  \bibfield  {author} {\bibinfo {author} {\bibfnamefont {S.~R.}\ \bibnamefont
  {White}},\ }\href {\doibase 10.1103/PhysRevB.48.10345} {\bibfield  {journal}
  {\bibinfo  {journal} {Phys. Rev. B}\ }\textbf {\bibinfo {volume} {48}},\
  \bibinfo {pages} {10345} (\bibinfo {year} {1993})}\BibitemShut {NoStop}%
\bibitem [{\citenamefont {Kumar}\ \emph {et~al.}(2010)\citenamefont {Kumar},
  \citenamefont {Soos}, \citenamefont {Sen},\ and\ \citenamefont
  {Ramasesha}}]{kumarmodifieddmrg}%
  \BibitemOpen
  \bibfield  {author} {\bibinfo {author} {\bibfnamefont {M.}~\bibnamefont
  {Kumar}}, \bibinfo {author} {\bibfnamefont {Z.~G.}\ \bibnamefont {Soos}},
  \bibinfo {author} {\bibfnamefont {D.}~\bibnamefont {Sen}}, \ and\ \bibinfo
  {author} {\bibfnamefont {S.}~\bibnamefont {Ramasesha}},\ }\href {\doibase
  10.1103/PhysRevB.81.104406} {\bibfield  {journal} {\bibinfo  {journal} {Phys.
  Rev. B}\ }\textbf {\bibinfo {volume} {81}},\ \bibinfo {pages} {104406}
  (\bibinfo {year} {2010})}\BibitemShut {NoStop}%
\end{thebibliography}%
\end{document}


\title{Supplemental Material: Emergent Quadrupolar Order in the Spin-1/2 Kitaev-Heisenberg Model}

\begin{abstract}
	We here provide supplemental explanations and data on the following topics in relation to the main text: (I) Schematic picture of quadrupolar ordered states, (II) Wave function analysis in two-leg ladder, (III) Quadrupolar order and correlation in 2-leg ladder from ED, (IV) Finite-size effects in the quadrupolar correlation functions in 2-leg ladder.
\end{abstract}
\author{Manodip Routh}
\thanks{These authors contributed equally to this work.}
\affiliation{S.N. Bose National Centre for Basic Sciences, Kolkata 700098, India.}
\author{Sayan Ghosh}
\thanks{These authors contributed equally to this work.}
\affiliation{S.N. Bose National Centre for Basic Sciences, Kolkata 700098, India.}
\author{Jeroen van den Brink}
\affiliation{Institute for Theoretical Solid State Physics, IFW Dresden, 01069 Dresden, Germany.}
\affiliation{\mbox{W\"urzburg-Dresden Cluster of Excellence ct.qmat, Germany}}
\affiliation{Department of Physics, Technical University Dresden, 01069 Dresden, Germany.}
\author{Satoshi Nishimoto}
\email{s.nishimoto@ifw-dresden.de}
\affiliation{Institute for Theoretical Solid State Physics, IFW Dresden, 01069 Dresden, Germany.}
\affiliation{Department of Physics, Technical University Dresden, 01069 Dresden, Germany.}
\author{Manoranjan Kumar}
\email{manoranjan.kumar@bose.res.in}
\affiliation{S.N. Bose National Centre for Basic Sciences, Kolkata 700098, India.}
\date{\today}
\maketitle

\section{\label{SM1}Schematic picture of quadrupolar states}
\begin{figure}[h!]
	\centering
	\includegraphics[width=0.8\textwidth]{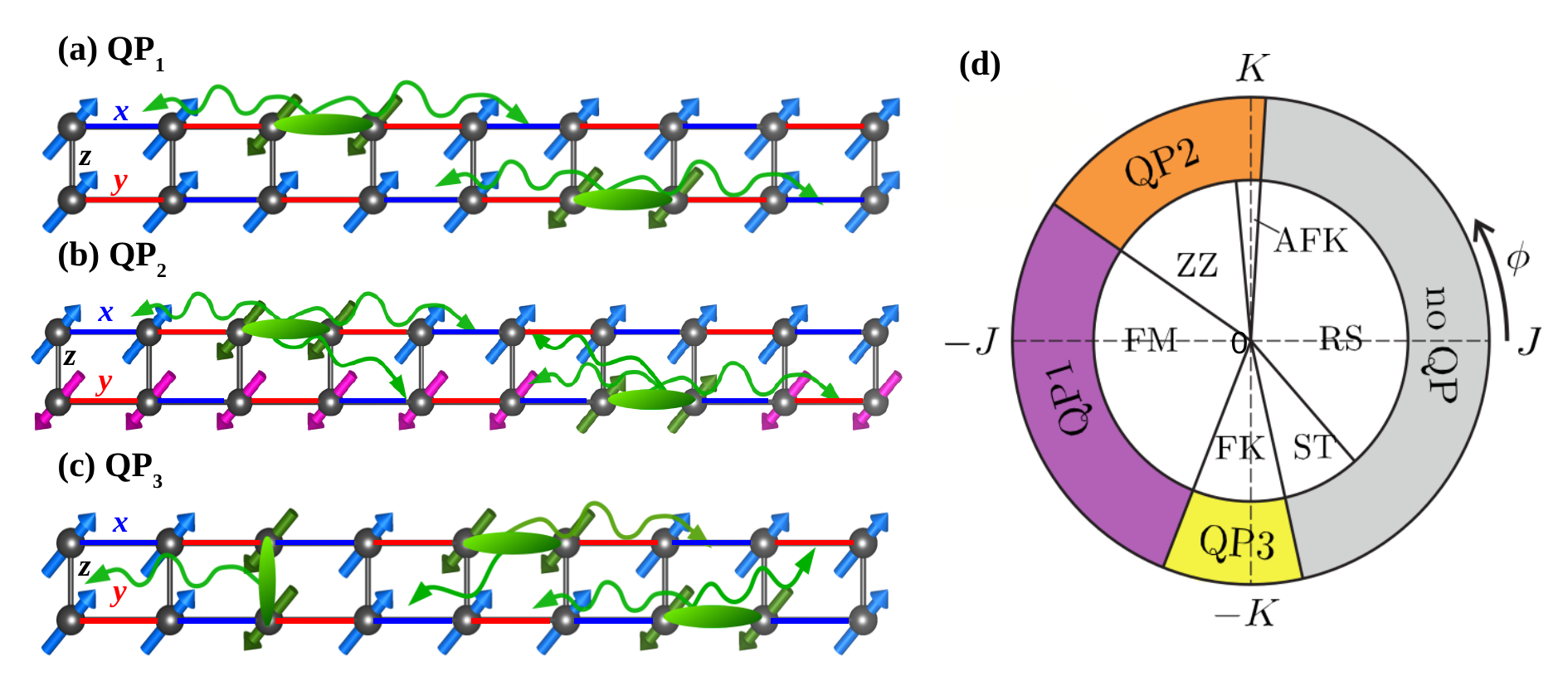}
	\caption{Spin nematic state expected for a KH ladder. Schematic representation of bound pairs of magnons moving in a background of polarized spins. Here, most of the spins colored as blue and magenta arrows align in $+z$ or $-z$ directions, respectively. Flipped spins (black arrows) gain energy by propagating as bound pairs along (a) same leg in QP$_1$ phase, (b) both the same leg and across the leg in QP$_2$ phase, (c) rung, same leg and across the leg in QP$_3$ phase. (d) Phase diagram for magnetic order (inner) and QP order (outer) (same as main text).}
\label{fig:QPschematic}
\end{figure}

We explain some details about the structure of quadrupolar (QP) order in the two-leg Kitaev-Heisenberg ladder. In Fig.\ref{fig:QPschematic} schematic picture of quadrupolar states is shown. Both $P_{\rm L1L2}(r)$ and $P_{\rm RR}(r)$ are always positive i.e. indicates non-staggered behaviour. The first type of phase involves the confinement of two spin flips along the leg, as depicted in Fig.\ref{fig:QPschematic}(a). This phase is denoted as QP$_1$ and exhibits long-range order along the leg ($P_{\rm L1L1}(r)$). The second type of phase is QP$_2$, featuring long-range pair correlation along the same leg ($P_{\rm L1L1}(r)$) and across the leg ($P_{\rm L1L2}(r)$), as illustrated in Fig. \ref{fig:QPschematic}(b). The third type is QP$_3$, exhibits long-range pair correlation along the rung ($P_{\rm RR}(r)$), the same leg ($P_{\rm L1L1}(r)$), and across the leg ($P_{\rm L1L2}(r)$), as shown in Fig. \ref{fig:QPschematic}(c).

\newpage
\section{\label{SM2}Wave function analysis in two-leg ladder}
\textit{Zigzag phase}$-$ The range of this phase is $\phi = 0.53\pi$ to $0.81\pi$, with exchange interactions $J$ and $K$ being ferromagnetic (FM) and antiferromagnetic (AFM), respectively. The spins on each leg are ferromagnetically aligned, but the spins on one leg are in the opposite direction to those on the other. The ground state properties of this phase can be analyzed using the wave function. The wave function at $\phi \approx 0.65\pi$ can be written as follows,

\begin{eqnarray}    
	\psi(\phi) &\approx& [(C_{1} + \sum_{m=1}^{N/2} C_{2}(-1)^m (S_{m,1}^{+}S_{m+1,1}^{+} + S_{m,2}^{-}S_{m+1,2}^{-})  + C_{3} S_{m,1}^{+}S_{m,2}^{-}) + \sum_{m\neq n=1}^{N/2}C_{4} S_{m,1}^{+}S_{m+1,1}^{+}S_{n,1}^{+}S_{n+1,1}^{+} + \cdots ]\prod_{i=1}^{N/2}S_{i,1}^{-}\ket{\Uparrow} \nonumber \\ 
	&+& [(C_{1} + \sum_{m=1}^{N/2} C_{2}(-1)^m (S_{m,1}^{-}S_{m+1,1}^{-}
	+ S_{m,2}^{+}S_{m+1,2}^{+}) + C_{3} S_{m,1}^{-}S_{m,2}^{+}) + \sum_{m\neq n=1}^{N/2}C_{4} S_{m,1}^{-}S_{m+1,1}^{-}S_{n,1}^{-}S_{n+1,1}^{-} + \cdots]\prod_{i=1}^{N/2}S_{i,1}^{+}\ket{\Downarrow} \nonumber \\
	&+& \cdots 
	\label{Eq:ZZ_WF}
\end{eqnarray}

Here, $\ket{\Uparrow}$ and $\ket{\Downarrow}$ denote spin-up and spin-down states, respectively. The largest coefficient $C_{1}$ corresponds to two configurations where spins in one leg align along the $+$z direction and spins in the other leg align along the $-$z direction. The next dominant configurations with coefficients $C_{2}$ and $C_{3}$ correspond to $2N$ and $N$ configurations, exhibiting a simultaneous double spin flip along the leg and the rung. $C_{4}$ corresponds to $2N$ configurations with a simultaneous four-spin flip along the leg. The contributions from other configurations involving multiple spin flips are comparatively small. The system size dependence of these coefficients is shown in the Table.~\ref{supp-table1}. Due to the large quantum fluctuations in this phase, all the configurations contribute substantially to the QP order, which accounts for the highest QP order in this phase.

\begin{table}[h]
\caption{Dominant coefficients of the ground state wave function in the zigzag phase at $\phi = 0.65\pi$ with $N=8 \times 2$ and $10 \times 2$ sites ladder.}
\label{supp-table1}
\centering
\begin{tabular}{|c|c|c|c|c|}
\hline
Coefficients & $N=8 \times 2$ &No. of diagrams & $N=10 \times 2$ & No. of diagrams \\
\hline
$C_1$ & $0.43387$ & $2$ & $0.34001$ & $2$ \\
$C_2$ & $0.09194$ & $32$ ($2N$) & $0.07170$ & $48$ ($2N$) \\
$C_3$ & $0.08103$ & $16$ ($N$) & $0.06424$ & $16$ ($N$)   \\
$C_4$ & $0.02541$ & $32$ ($2N$) & $0.01947$ & $48$ ($2N$) \\
\hline
\end{tabular}
\end{table}

\textit{Ferro-Kitaev phase}$-$ The FK phase occurs in the range $1.37\pi < \phi < 1.56\pi$. The gs wave function at $\phi \approx 1.53\pi$ can be approximated as,
\begin{eqnarray}
	\psi(\phi) &\approx& [C_{1} + \sum_{m=n=1}^{N/2} C_{2} S_{m,1}^{-}S_{m+1,1}^{-}S_{n,2}^{-}S_{n+1,2}^{-} 
	+ S_{m,1}^{-}S_{m,2}^{-}S_{n,1}^{-}S_{n,2}^{-}]  \ket{\Uparrow} \nonumber \\
	&+& [C_{1} + \sum_{m=n=1}^{N/2} C_{2} S_{m,1}^{+}S_{m+1,1}^{+}S_{n,2}^{+}S_{n+1,2}^{+}  
	+ S_{m,1}^{+}S_{m,2}^{+}S_{n,1}^{+}S_{n,2}^{+}  \ket{\Downarrow}] +\cdots
	\label{Eq:FK_WF}
\end{eqnarray}
Here, double spin flips occur only along the rung. $C_1$ corresponds to configurations where all the spins align along $+$z or $-$z direction. $C_2$ corresponds to simultaneous double-spin flips along rung. Due to strong Kitaev interaction, quantum fluctuations are very high in this phase, and coefficients are comparable. The other configurations with multiple spin flips are with small contributions. These configurations contribute significantly to the QP order parameter, which is finite in this phase.
\newpage

\section{\label{SM3}Quadrupolar order and correlation in 2-leg ladder from ED}

\begin{figure}[th]
	\centering
	\begin{subfigure}[b]{0.40\textwidth}
		\centering
		\includegraphics[width=\textwidth]{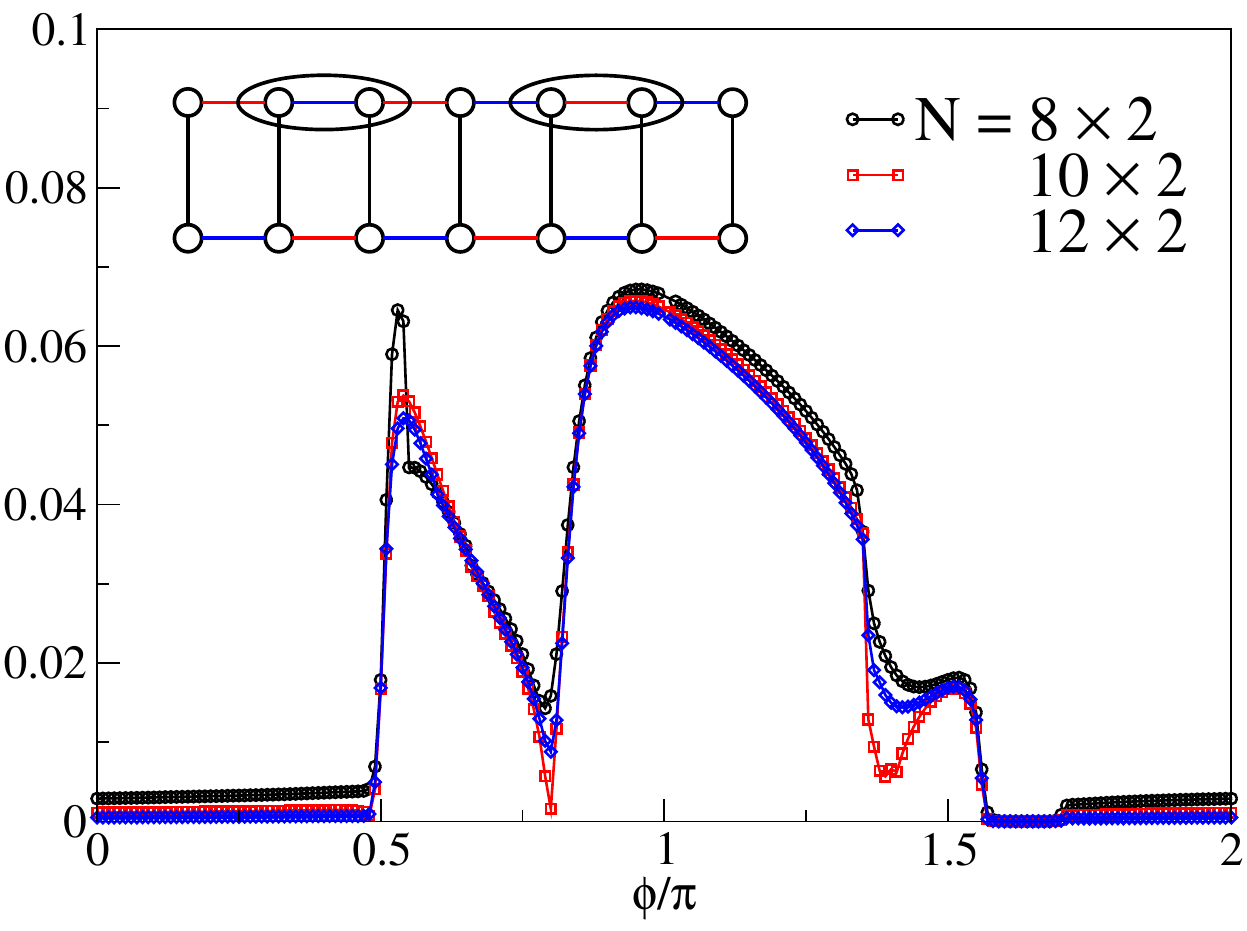}
		\caption{}
		\label{fig:quadcor_sameleg}
	\end{subfigure}
	\hfill
	\begin{subfigure}[b]{0.40\textwidth}
		\centering
		\includegraphics[width=\textwidth]{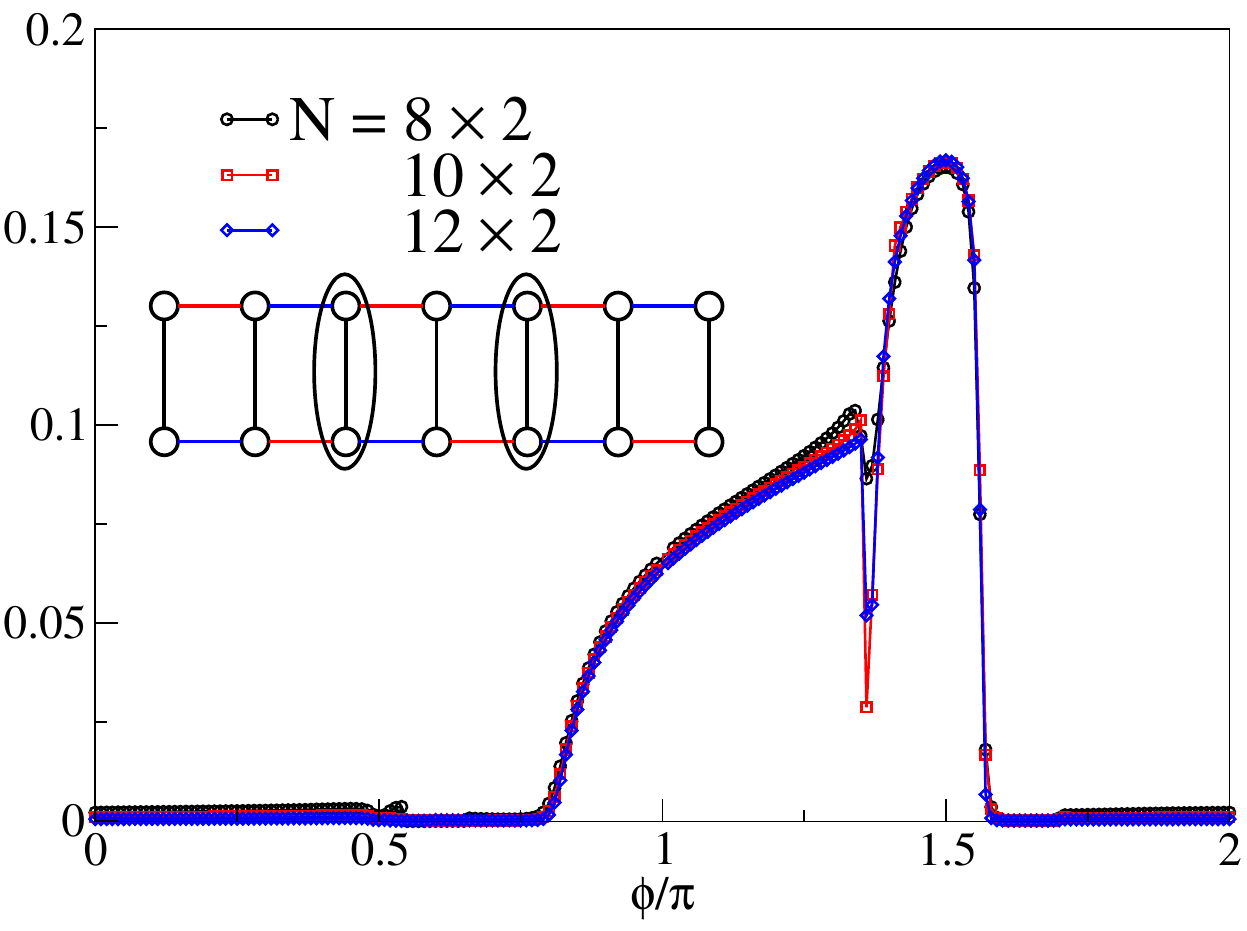}
		\caption{}
		\label{fig:quadcor_rung}
	\end{subfigure}
	\hfill
	\begin{subfigure}[b]{0.40\textwidth}
		\centering
		\includegraphics[width=\textwidth]{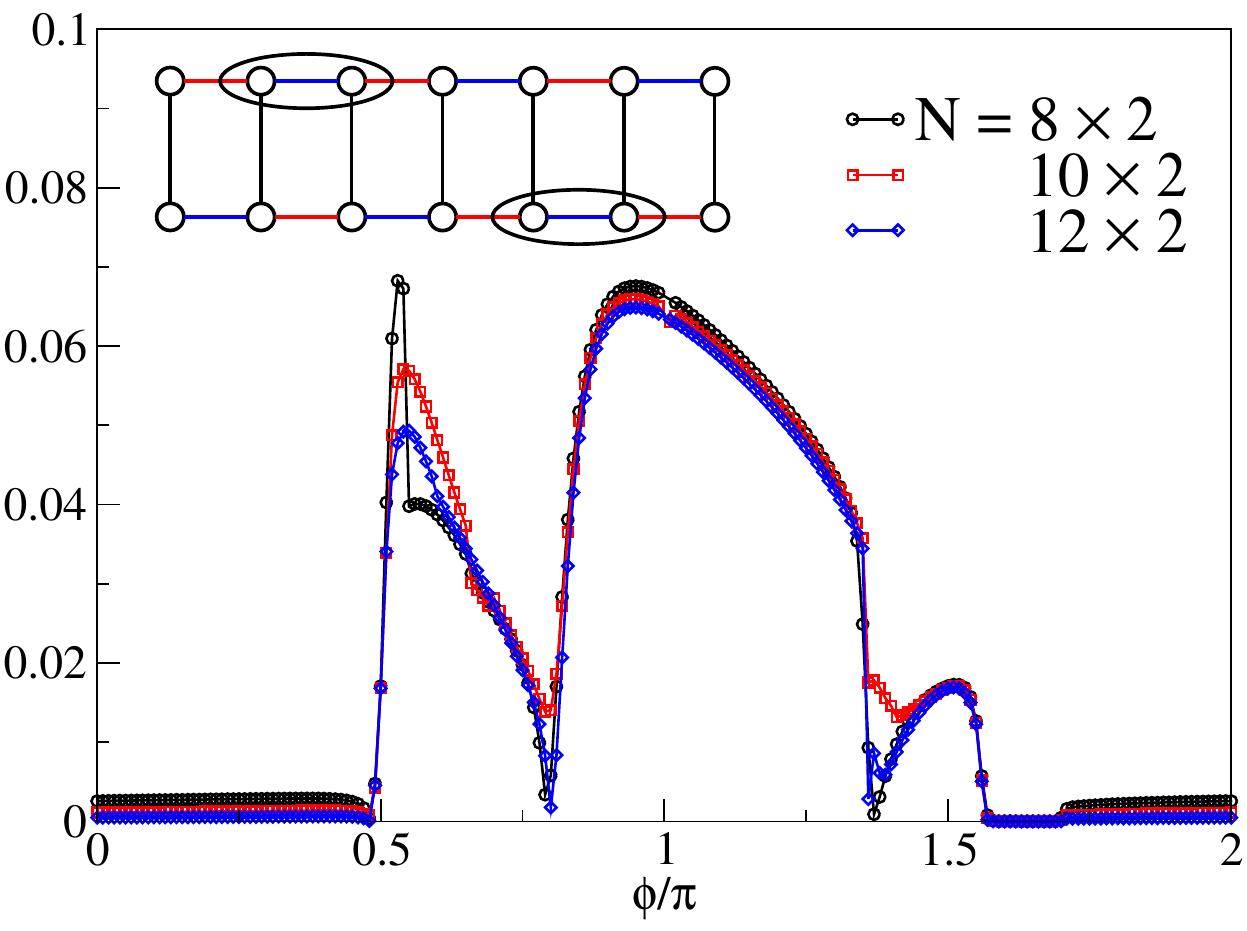}
		\caption{}
		\label{fig:quadcor_diffleg}
	\end{subfigure}	
	\caption{QP correlation between two most distant bonds in the (a) same leg, (b) rung and (c) different legs for $N = 8 \times 2$, $10 \times 2$ and $12 \times 2$ sites ladder with periodic boundary conditions along both the direction(leg and rung). The schematic view of the ladder with the two most distant bonds indicated by the ellipse is shown in the inset.}
	\label{fig:quadcor_all}
\end{figure}


The QP correlation $|\langle S_{i}^{+}S_{j}^{+}S_{i+r}^{-}S_{j+r}^{-}\rangle |$ between two most distant bonds is calculated in various ways for $N = 8 \times 2$, $10 \times 2$ and $12 \times 2$ sites ladder with periodic boundary conditions along both the direction(leg and rung). Fig.~\ref{fig:quadcor_sameleg} depicts the correlation between the two most distant bonds on the same leg. Fig.~\ref{fig:quadcor_rung} depicts the correlation between the two most distant rung bonds. Fig.~\ref{fig:quadcor_diffleg} depicts the correlation between the two most distant bonds in different legs. These correlations depend on the system size ($N$) and will disappear in the thermodynamic limit.

\newpage

\section{\label{SM4}Finite-size effects in the quadrupolar correlation functions in 2-leg ladder}
We apply the density-matrix renormalization group (DMRG) technique to calculate the quadrupolar correlation for larger system size. DMRG is a state-of-the-art numerical technique to calculate accurate ground states and a few low-lying excited energy states of strongly interacting quantum systems \cite{white1992,white1993}. It is based on the systematic truncation of irrelevant degrees of freedom. We use a modified DMRG scheme \cite{kumarmodifieddmrg} with open boundary conditions where two sites are added in each block at every step. The system starts with $4$ spins and grows to $N = 4n$ in $N - 1$ steps. The Fock space dimensionality of each block increases as $2(2s + 1)m$ or as $4m$ for $s = 1/2$ system. Keeping up to $m = 5000 $ eigenvectors of the density matrix is sufficient for good accuracy. The truncation error in the sum of the eigenvalues of the density matrix is less than $2 \times 10^{-5}$ in the worst case, and increasing $m$ changes the energy only in the forth or fifth decimal place. 

\begin{figure}[h!]
	\centering
	\begin{subfigure}[b]{0.45\textwidth}
		\centering
		\includegraphics[width=\textwidth]{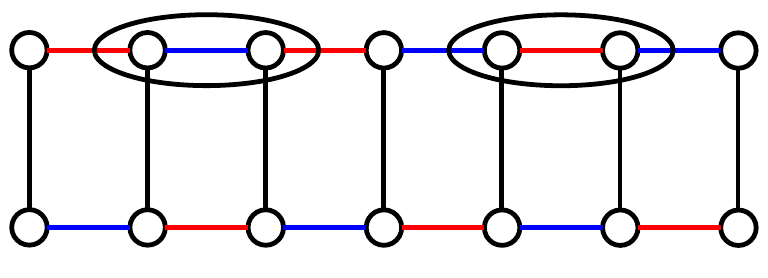}
		\caption{}
		\label{fig:schem_sameleg}
	\end{subfigure}
	\hfill
	\begin{subfigure}[b]{0.45\textwidth}
		\centering
		\includegraphics[width=\textwidth]{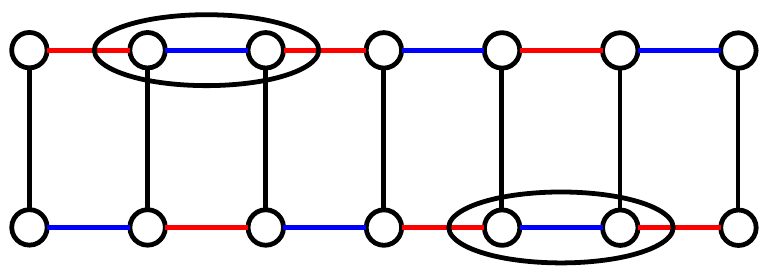}
		\caption{}
		\label{fig:schem_diffleg}
	\end{subfigure}
	\hfill
	\begin{subfigure}[b]{0.45\textwidth}
		\centering
		\includegraphics[width=\textwidth]{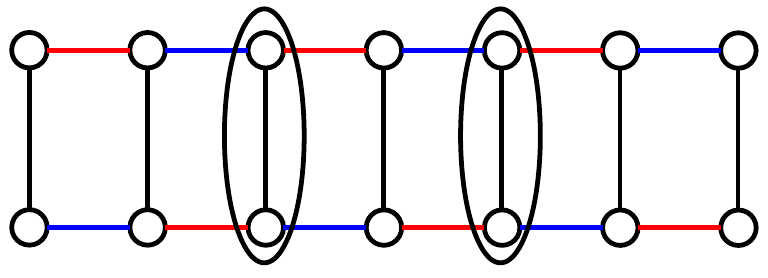}
		\caption{}
		\label{fig:schem_rung}
	\end{subfigure}	
	\caption{Schematic structure of a two-leg ladder. (a) $P_{\rm L1L1}(r)$ is calculated between bonds in the same leg indicated by an ellipse. (b) $P_{\rm L1L2}(r)$ is calculated between bonds in the different legs indicated by an ellipse. c) $P_{\rm RR}(r)$ is calculated between rung bonds indicated by ellipse.}
\label{fig:quadcor_all}
\end{figure}

In the main text, we argue the emergence of long-range QP order for a wide range of $\phi$ based on the results of quadrupolar correlation functions. We here investigate the finite-size effects appearing in the quadrupolar correlation functions. For the case of 2-leg ladder we calculate the following three kinds of quadrupolar correlation functions:

\noindent
(1) The quadrupolar correlation between bonds in the same leg is defined as
\begin{equation}
P_{\rm L1L1}(r) = \langle S_{i,1}^+ S_{i+1,1}^+ S_{i+r,1}^- S_{i+1+r,1}^- \rangle.
\label{Eq1}
\end{equation}
(2) The correlation between bonds the different legs is defined as
\begin{equation}
P_{\rm L1L2}(r) = \langle S_{i,1}^+ S_{i+1,1}^+ S_{i+r,2}^- S_{i+1+r,2}^- \rangle.
\label{Eq2}
\end{equation}
(3) The correlation between rung bonds is defined as
\begin{equation}
P_{\rm RR}(r) = \langle S_{i,1}^+ S_{i,2}^+ S_{i+r,1}^- S_{i+r,2}^- \rangle.
\label{Eq3}
\end{equation}
In Fig.~\ref{fig:quadcor_all} schematic ladders are presented to show how the correlations are calculated. $P_{\rm L1L1}(r)$ is calculated between bonds in the same leg indicated by an ellipse in Fig.~\ref{fig:schem_sameleg}. $P_{\rm L1L2}(r)$ is calculated between bonds in different legs indicated by an ellipse in Fig.~\ref{fig:schem_diffleg}. $P_{\rm RR}(r)$ is calculated between bonds in the rung indicated by an ellipse in Fig.~\ref{fig:schem_rung}.

\begin{figure}[h!]
	\centering
	\includegraphics[width=0.6\textwidth]{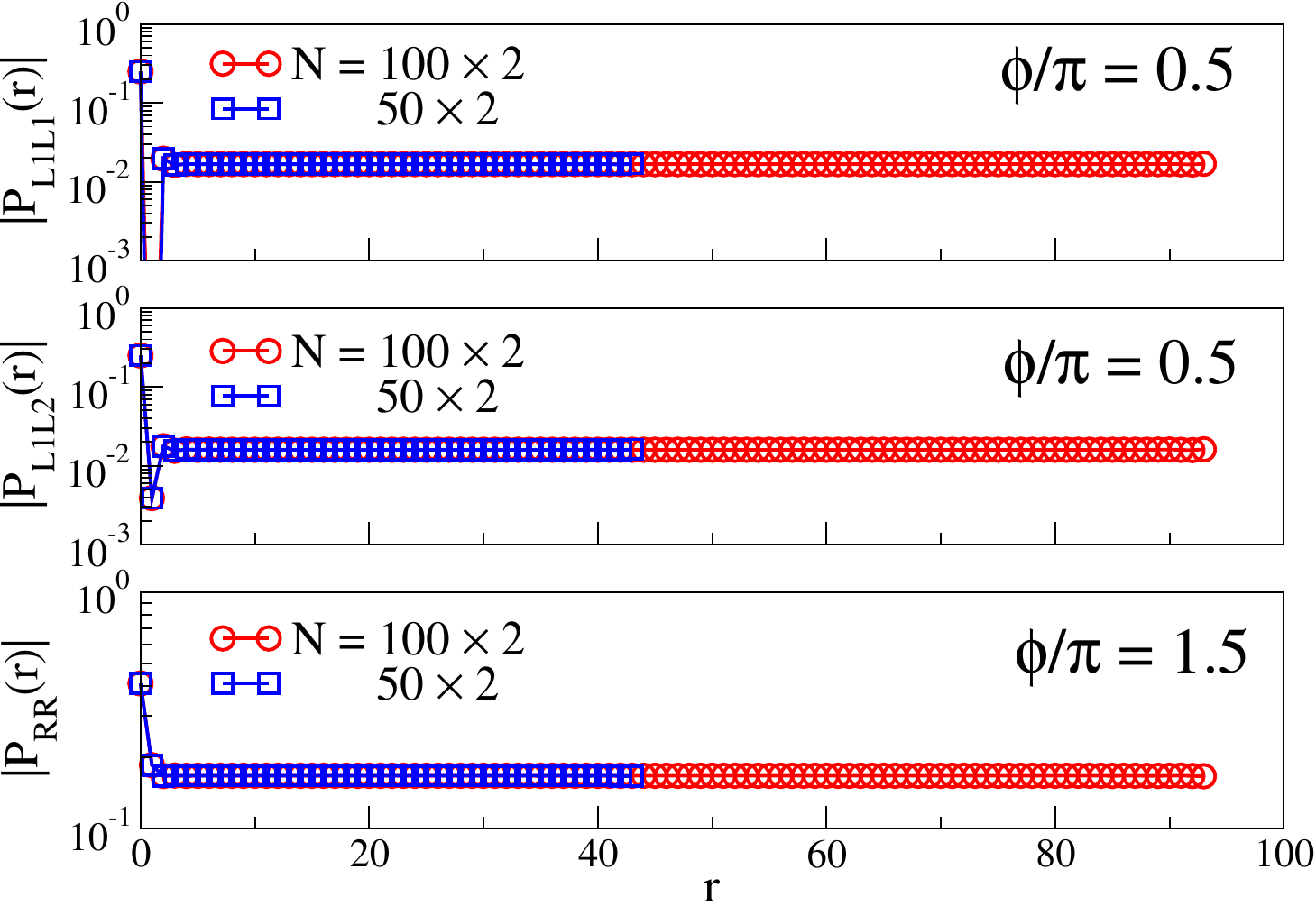}
	\caption{Size dependence of various quadrupolar correlation functions $P_{\rm L1L1}(r)$, $P_{\rm L1L2}(r)$, and $P_{\rm RR}(r)$ at $\phi=0.5\pi$, $1.5\pi$ for $N=50 \times2 $ and $100 \times 2$ site ladders.}
	\label{fig:size}
\end{figure}

Fig.~\ref{fig:size} we plot the DMRG results for various quadrupolar correlation functions $P_{\rm L1L1}(r)$, $P_{\rm L1L2}(r)$, and $P_{\rm RR}(r)$ at $\phi=0.5\pi$ and $1.5\pi$ using $N=50 \times 2$ and $100 \times 2$ site ladders. In these cases, the correlation functions exhibit long-range QP order. For all cases the $N=100 \times 2$ results appear to be smoothly extend the $N=50 \times 2$ ones to larger distance. Therefore, we can confirm the almost negligible system-size dependence.  

\bibliography{KH_ref}